\documentclass[11pt]{article}
\usepackage{geometry}                
\usepackage{supertabular}
\usepackage{subfigure}
\usepackage{longtable}
\usepackage{color}

\usepackage{graphicx}
\usepackage{amssymb}
\usepackage{amsmath}
\usepackage{epstopdf}
\usepackage{multicol}
\usepackage[square, comma, compress]{natbib}

\title{Axino dark matter and baryon number asymmetry production\\ by the $Q$-ball decay in gauge mediation}
\author{Shinta Kasuya$^a$, Etsuko Kawakami$^b$ and Masahiro Kawasaki$^{b,c}$\\
\\
$^a$ \small{Department of Mathematics and Physics,
     Kanagawa University,  Kanagawa 259-1293, Japan} \\
$^b$ \small{Institute for Cosmic Ray Research,
     University of Tokyo, Chiba 277-8582, Japan}\\
$^c$ \small{Kavli Institute for the Physics and Mathematics of the Universe (WPI),}\\
     \small{Todai Institutes for Advanced Study, University of Tokyo, Chiba 277-8582, Japan}}
\date{November 18, 2015}                   
\begin{document}

\maketitle

\begin{abstract}
We investigate the $Q$-ball decay into the axino dark matter in the gauge-mediated 
supersymmetry breaking. In our scenario, the $Q$ ball decays mainly into nucleons 
and partially into axinos to account respectively for the baryon asymmetry and the dark 
matter of the universe. The $Q$ ball decays well before the big bang 
nucleosynthesis so that it is not affected by the decay. We show the region of the 
parameters which realizes this scenario.
\end{abstract}

\section{Introduction}

The origins of baryon number asymmetry and dark matter of the universe have been 
discussed for decades, but are still some of the main unsolved mysteries in cosmology. 
In supersymmetry (SUSY), the lightest 
supersymmetric particle (LSP), with $R$-parity conservation, is stable and, in most 
cases, scarcely interacts with other particles. These natures make the LSP a strong 
candidate of dark matter. SUSY could not only give a candidate of the 
dark matter, but may also explain the origin of baryon number asymmetry. 
In the Minimal Supersymmetric Standard Model (MSSM), 
there exist many flat directions, which consist of squarks and sleptons (and the Higgs fields in some cases), 
thus carrying baryon and/or lepton numbers. Therefore, the flat direction could be responsible 
for baryogenesis through the Affleck-Dine (AD) mechanism \cite{AD} and is called the AD field. 
The AD field, carrying the baryon number, has a large VEV during inflation. It begins rotation 
in the potential after inflation, and the baryon number is created. It finally decays into quarks 
to become the baryon asymmetry of the universe. 

The very attractive feature of the AD mechanism is to provide both the baryon asymmetry 
and dark matter of the universe simultaneously in the context of the $Q$-ball cosmology 
\cite{KuSh,EnMc,KK1,KK2,New,KK3,FuHa,Seto,RS,ShKu,DoMc,KK2011,KKY2013,KK2014a,KK2014b}.  
During the rotation, the AD condensate may fragment into non-topological solitons, $Q$ balls. 
These $Q$ balls can be dark matter if they are stable, while the LSP dark matter could be
produced from unstable $Q$ balls. Stable $Q$ balls form if the charge $Q$ is large
enough in the gauge-mediated SUSY breaking \cite{KuSh,New,KK3}. 
On the other hand, $Q$ balls are unstable in the gravity mediation producing the
neutralino LSP \cite{EnMc,KK2,FuHa}, the gravitino LSP \cite{Seto}, and the axino LSP \cite{RS},
and in the gauge mediation creating the gravitino LSP if the charge is small enough 
\cite{ShKu,DoMc,KK2011,KKY2013}.

In this paper, we investigate a model that the $Q$ ball decays into axino LSPs in 
gauge-mediated SUSY breaking. The axino is a fermionic superpartner of the axion, 
which is introduced as a dynamical scalar field to solve the strong CP problem in 
quantum chromodynamics known as Peccei-Quinn (PQ) mechanism \cite{PQ}. 
In our model, the $Q$ ball decays mainly into 
nucleons and partially into axinos directly in order to account for both the baryon 
asymmetry and dark matter of the universe. The decay of $Q$ ball takes place 
well before the big bang nucleosynthesis (BBN) so that the decay itself does not 
affect the BBN. The $Q$-ball decay could produce the lightest supersymmetric particle (LSP) of 
the MSSM, whose decay may destroy light elements synthesized during the BBN. 
However, the MSSM LSPs (MLSPs) would annihilate afterwards \cite{DM,KKY2013}, 
and the resultant abundance of the MLSPs is small enough to avoid the serious BBN constraints
typically for $m_{3/2} \lesssim$~GeV. 

$Q$ balls may also decay into gravitinos in our scenario \cite{KK2011,KKY2013}. 
In most region of the parameter space, the branching into the gravitino is 
much smaller than that of the axino because of the much stronger coupling to the axino
than to the gravitino, and we may well neglect the contribution of the gravitino abundance 
to the dark matter density. Also notice that we assume the axion density does not contribute
to the dark matter density in our scenario. This is simply achieved by setting the misalignment 
angle small enough. 

The structure of this paper is as follows. After briefly reviewing the $Q$-ball features in the 
gauge mediation in the next section, we show the details of the decay process of the $Q$ 
ball in Sec.3. In Sec.4, we obtain the baryon and the axino dark matter abundances as well 
as the MLSP abundance. In Sec.5, we show the realization of those successful scenario in the 
$Q$-ball parameters, taking also into account the constraints on the MLSP abundance by 
the BBN. Finally, we summarize our results in Sec.6.  Appendices are devoted to some details 
of the axino productions in the SUSY axion models which are used in the main text.

\section{$Q$ ball in gauge mediation}
The AD field $\Phi$ is a combination of the squarks, the sleptons and the Higgs whose potential is flat in 
the SUSY exact limit. Because of the SUSY breaking in the gauge mediation, the flat potential 
is lifted such that $V \sim m_{\phi}^2 \phi^2$ below the messenger scale, while it is flat 
above the messenger scale, $V\sim M^4_{F}$ \cite{log2,KuSh}. Here $m_{\phi}$ is a 
soft SUSY breaking mass and $M_{F}$ is related to the $F$ component of a 
gauge-singlet chiral multiplet $S$ in the messenger sector as 
$M_{F}^4 \equiv \frac{g^2}{(4\pi)^4}\langle F_{S}\rangle^2$ where $g$ is a gauge coupling 
constant in the standard model, and $M_{F}$ is allowed in the following range:
\begin{equation}
\label{Mf}
4 \times 10^4\ \mathrm{GeV} \lesssim M_{F} \lesssim \frac{g^{1/2}}{4\pi}\sqrt{m_{3/2}M_{P}},
\end{equation}
where $m_{3/2}$ and $M_{P}=2.4 \times 10^{18}\ \rm GeV$ are the gravitino and the reduced
Planck masses, respectively.

When the Hubble parameter becomes smaller 
than the curvature of the potential, the AD field begins to oscillate and the baryon number is created. 
During the helical motion, it transforms into $Q$ balls. The typical charge of the 
formed $Q$ ball is estimated as \cite{KK1}
\begin{equation}
Q = \beta \left( \frac{\phi_{\mathrm{osc}}}{M_{F}}\right)^{4},
\end{equation}
where $\phi_{\mathrm{osc}}$ is the field amplitude when the oscillation begins, and 
$\beta \simeq 6\times 10^{-4}$ when the oscillating field has a nearly circular orbit 
$\epsilon = 1$ ($\epsilon$: ellipticity of the orbit) and $\beta \simeq 6\times 10^{-5}$ 
when $\epsilon \lesssim 0.1$. The charge $Q$ is just the $\Phi$-number, and relates 
to the baryon number of the $Q$ ball as
\begin{equation}
B = bQ,
\end{equation}
where $b$ is the baryon number carried by a $\Phi$ particle. For example, 
$b=\frac{1}{3}$ for the \textit{udd} direction. The mass, the size, the rotation velocity 
and the field value at the center of the $Q$ ball are related to the charge $Q$ as
\begin{eqnarray}
\label{Qballpropaties}
M_{Q} \simeq \frac{4\sqrt{2}\pi}{3}\zeta M_{F}Q^{3/4}, \label{MQ}\\
R_{Q} \simeq \frac{1}{\sqrt{2}}\zeta^{-1}M_{F}^{-1}Q^{1/4}, \\
\omega_{Q} \simeq \sqrt{2}\pi \zeta M_{F}Q^{-1/4}, \\
\phi_{Q} \simeq \frac{1}{\sqrt{2}}\zeta M_{F}Q^{1/4}, 
\end{eqnarray}
respectively. Here $\zeta$ is the $O$(1) parameter determined by the fit to numerical calculation 
\cite{KKY2013}, and we adopt $\zeta \approx 2.5$.

\section{$Q$-ball decay}
A $Q$-ball decay occurs when some decay particles have the same kind of charges 
as the $Q$ ball and the mass of each decay particle is less than $\omega_Q$. 
Here we are interested in the case where the $Q$ ball decays into the quarks but not into MLSPs. 
It is described by the condition $bm_{\mathrm{N}} < \omega_{Q} < m_{\mathrm{MLSP}}$ 
where $m_{N}$ and $m_{\mathrm{MLSP}}$ are the nucleon and MLSP masses, respectively. 
This implies that the $Q$-ball charge should be $Q_{\mathrm{cr}} < Q < Q_{\rm D}$ where
\begin{eqnarray}
& & Q_{\mathrm{cr}}  = 4\pi^4\zeta^4 \left( \frac{M_{F}}{m_{\mathrm{MLSP}}} \right)^4, \\
& & Q_{\rm D}  =   4\pi^4\zeta^4 \left( \frac{M_{F}}{bm_{N}}\right)^4.
\label{Q2N}
\end{eqnarray}
If the $Q$ ball decays into quarks, the Pauli blocking effects suppress the rate.
Thus the decay rate $\Gamma_Q$ has an upper bound $\Gamma_Q^{\mathrm{(sat,d)}}$ 
corresponding to the maximum flux of the quarks from the surface of the $Q$-ball \cite{Cohen}. 
For the decay into two distinguishable particles, the upper bound is given by \cite{KKY2013},
\begin{equation}
\label{satdecayrate}
\Gamma_{Q} \lesssim \Gamma_Q^{\rm (sat, d)}
\simeq \frac{1}{Q} \frac{\omega_{Q}^3}{96\pi^2} 4\pi R_{Q}^2 
\simeq \frac{\pi^2}{12\sqrt{2}} M_{F} Q^{-5/4}\zeta.
\end{equation}
This saturation occurs approximately for $f_{\mathrm{eff}}\phi_Q \gtrsim \omega_{Q}$, 
where  $f_{\mathrm{eff}}$ is the effective coupling constant by which the interaction is 
written as $\mathcal{L}_{\mathrm{int}} = f_{\mathrm{eff}} \phi^{*}\chi\eta + \mathrm{h.c}$, 
where $\chi$ and $\eta$ are the particles that the $Q$-ball decays into.

The elementary process of the $Q$-ball decay into nucleon is 
squark + squark $\rightarrow$ quark + quark via gluino exchanges for $\omega_Q < m_{\tilde{g}}$,
where $m_{\tilde{g}}$ is the gluino mass.
The effective coupling constant $f_{\rm eff}$ of this process for $\phi_{Q} > m_{\tilde{g}}$ is given by 
$f_{\mathrm{eff}} \simeq \frac{m_{\tilde{g}}}{\phi_Q}$ \cite{Kawasaki:2012gk},
Since we suppose $m_{\tilde{g}} \simeq 1 \,\mathrm{TeV} > m_{\mathrm{MLSP}}$, we have
\begin{equation}
\label{Qballdecay}
\frac{f_{\mathrm{eff}}\phi_{Q}}{\omega_{Q}} 
=\frac{m_{\tilde{g}}}{\omega_Q} > 1,
\end{equation}
we can see that this process is saturated. The decay rate of the $Q$-ball into quarks is given 
by \cite{Kawasaki:2012gk}
\begin{equation}
\Gamma^{\rm (q)}_{Q} = 1.1 \times 8N_{q}\Gamma^{\rm (sat, d)}
\end{equation}
The factor 8 comes from the fact that $\omega_{Q}$ in Eq.(\ref{satdecayrate}) should be replaced by 
$2\omega_{Q}$, since this process involves a decay of two squarks. $N_{q}$ is the possible degrees
of freedom of the quarks and here we set $N_q = 3\times3\times2=18$. Therefore, $Q$ balls decay 
at the cosmic time $t \simeq 1/\Gamma^{q}_{Q}$ when the universe is radiation-dominated.
The cosmic temperature at the $Q$-ball decay is estimated as
\begin{eqnarray}
\label{td}
T_{\mathrm{D}} &\simeq& 
\left( \frac{90}{4\pi^2 N_{d}}\right)^{1/4}\sqrt{\Gamma^{(\mathrm{q})}_{Q}M_{\mathrm{P}}},
\nonumber \\
& \simeq & 67 \,\mathrm{MeV} 
\left( \frac{M_{F}}{10^6\,\mathrm{GeV}}\right)^{1/2} 
\left( \frac{Q}{10^{23}}\right)^{-5/8} \left(\frac{N_{q}}{18}\right)^{1/2}
\left(\frac{N_{d}}{10.75}\right)^{-1/4}
\left( \frac{\zeta}{2.5}\right)^{1/2} ,
\end{eqnarray}
where $N_d$ is the relativistic degrees of freedom at $T_{\mathrm{D}}$. 
Note that if $Q$-ball charge is less than \cite{KK3}
\begin{equation}
Q_{\mathrm{evap}} \simeq 2.2 \times 10^{16} \left( \frac{M_{F}}{10^6\,\mathrm{GeV}}\right)^{-4/11}
\left(\frac{m_{\phi}}{\mathrm{TeV}}\right)^{-8/11},
\end{equation}
the $Q$ ball evaporates in thermal bath, but $Q$-ball charge in the allowed region of the successful 
scenario in this paper is large enough to survive from its evaporation.

Next, we consider the $Q$-ball decay into the axinos. The condition for the decay into 
axinos is described by $m_{\tilde{a}} < \omega_{Q}$. Using Eq.(\ref{MQ}), this 
condition is rewritten as
\begin{equation}
\label{malimit}
m_{\tilde{a}} < 20 \,{\rm GeV}  \left(\frac{M_{F}}{10^6\,{\rm GeV}}\right)
\left(\frac{Q}{10^{23}}\right)^{-1/4}\left( \frac{\zeta}{2.5}\right).
\end{equation}
The elementary process of the $Q$-ball decay into the axino is squark $\rightarrow$ quark + axino. 
The dominant part of the coupling comes from the logarithmically divergent part of the 
gluon-gluino-(s)quark loop term. The effective coupling is given by \cite{Covi:2002vw}
\begin{equation}
f_{\rm eff}^{(\tilde{a})} =  \frac{\alpha_{s}^2}{\sqrt{2}\pi^2} \frac{m_{\tilde{g}}}{f_a} 
\log{\left( \frac{f_a}{m_{\tilde{g}}}\right)}.
\end{equation}
In the DFSZ model, there also exists a tree-level axino-quark-squark coupling, but the
rate is proportional to (quark mass)$^2$ \cite{Choi:2011yf}, which is negligible in our 
scenario.
We thus obtain
\begin{equation}
\frac{f_{\rm eff}^{(\tilde{a})}\phi_Q}{\omega_Q} 
\simeq 3.6 \times10^{-2} \left( \frac{f_a}{10^{12} \, \mathrm{GeV}}\right)^{-1} 
\log{\left( \frac{f_a}{10^3 \,\mathrm{GeV}}\right)}\left( \frac{Q}{10^{23}}\right)^{1/2},
\end{equation}
where $f_a$ is the axion decay constant.\footnote{
We regard $f_a$ as $f_a/N_{\mathrm{c}}$ throughout the paper, 
where $N_{\mathrm{c}}$ is the color anomaly of the PQ symmetry.}
Here we take the coupling strength for strong
interaction as $\alpha_s = 0.1$ and the gluino mass $m_{\tilde{g}} = 1 \,\mathrm{TeV}.$
It depends on parameters $f_{a}$ and $Q$ whether the decay will be saturated. This is contrasted 
to the case that the decay into gravitino is not saturated \cite{KK2011}. 

The actual saturation is not determined simply by the condition $f_{\rm eff}^{(\tilde{a})}\phi_Q>\omega_Q$. 
We also have to consider the Pauli blocking effects of the quarks produced by the main channel of the
squark decay via $\tilde{q}+\tilde{q}\rightarrow q+q$ \cite{Kawasaki:2012gk}. Taking this effect into account, 
the branching ratio for the decay into axino is saturated for $f_{\rm eff}^{(a)} > f_{\rm eff}$, and estimated as
\begin{equation}
\label{branchsat}
B_{\tilde{a}}^{(\rm{sat})} 
= \frac{\Gamma^{(\mathrm{sat})}_{(\tilde{a})}}{\Gamma^{(q)}_{Q}} 
= \frac{1.4\Gamma^{(\mathrm{sat,d})}}{1.1\times 8N_q \Gamma^{(\mathrm{sat,d})}}
= 8.8 \times 10^{-3} \left(\frac{N_q}{18}\right)^{-1}. 
\end{equation}
On the other hand, the decay into axino is suppressed by the Pauli blocking effect for 
$f_{\rm eff}^{(a)} < f_{\rm eff}$. We denote the branching ratio in this case as 
$B_{\tilde{a}}^{(\rm{unsat})}$, calculated as 
\begin{eqnarray}
\label{branchunsat}
B_{\tilde{a}}^{(\rm{unsat})}
& \simeq & \left(\frac{f^{(\tilde{a})}_{\mathrm{eff}}}{f_{\mathrm{eff}}}\right)^2 
\simeq  \left( f_{\mathrm{eff}}^{(\tilde{a})} \frac{\phi_{Q}}{m_{\tilde{g}}}\right)^2 \nonumber \\
& = & 5.1\times10^{-7} \left( \frac{f_a}{10^{12}\,\mathrm{GeV}}\right)^{-2}
\left(\log{\frac{f_a}{10^3\,\mathrm{GeV}}}\right)^2 
\left(\frac{M_F}{10^6\,\mathrm{GeV}}\right)^2\left(\frac{Q}{10^{23}}\right)^{1/2}\left(\frac{\zeta}{2.5}\right)^2.
\end{eqnarray}

One may wonder if the gravitinos are abundantly produced by the $Q$-ball decay in 
this senario. To this end, we estimate the ratio $B_{\tilde{a}}/B_{3/2}$ :
\begin{eqnarray}
\label{branchingaxinotogravitino}
\frac{B_{\tilde{a}}}{B_{3/2}} \simeq 
\left\{
\begin{array}{ll}
\displaystyle{ 1.0 \times 10^{8}
\left(\frac{m_{3/2}}{10\,\rm MeV}\right)^2\left(\frac{M_F}{10^6\,{\rm GeV}}\right)^{-6}
\left(\frac{Q}{10^{23}}\right)^{1/2}\left(\frac{N_q}{18}\right)^{-1}\left( \frac{\zeta}{2.5}\right)^{-6} 
\left(\frac{m_{\tilde{g}}}{1\, \mathrm{TeV}}\right)^{2}}& \\
\hspace{50mm} \mathrm{(for\ the\ saturated\ case),}\\[5mm]
\displaystyle{ 5.8\times 10^3
\left( \frac{m_{3/2}}{10\,\rm MeV}\right)^2}\left( \frac{f_a}{10^{12}\, \mathrm{GeV}}\right)^{-2} 
 \left(\log{\frac{f_a}{10^3 \,\mathrm{GeV}}}\right)^2  & \\
\hspace{20mm}\displaystyle{\times \left( \frac{M_{F}}{10^6\,\mathrm{GeV}}\right)^{-4} 
\left( \frac{Q}{10^{23}}\right)
\left( \frac{\zeta}{2.5}\right)^{-4}\left(\frac{m_{\tilde{g}}}{1\, \mathrm{TeV}}\right)^2} \\
\hspace{50mm}\mathrm{(for\ the\ unsaturated\ case).}
\end{array}
\right.
\end{eqnarray}
Here we use \cite{Kawasaki:2012gk}
\begin{equation}
B_{3/2} \simeq \left(\frac{f_{3/2}}{f_{\rm eff}}\right)^2 
\simeq \left(\frac{\omega_{Q}^2}{\sqrt{3} m_{3/2}M_{P}}\frac{\phi_{Q}}{m_{\tilde{g}}}\right)^2.
\end{equation}
We simply consider the parameter space which satisfies $\frac{B_{\tilde{a}}}{B_{3/2}} > 1$ 
so that there is essentially no gravitino production in the $Q$-ball decay compared to the axino production. 

$Q$ balls also decay into MLSPs ($\chi$) when $\omega_Q$ becomes larger than $m_{\rm MLSP}$. 
The elementary process of the MLSP production is 
$\phi \rightarrow q + \chi$. Since $f_{\rm MLSP} \phi_Q / \omega_Q \gg 1$ and 
$f_{\rm MLSP} > f_{\rm eff}$, where $f_{\rm MLSP} \sim g$, the decay is saturated and the branching 
ratio is estimated as \cite{KKY2013}
\begin{equation}
\label{branchMLSP}
B_{\rm MLSP}
= \frac{\Gamma^{\rm (MLSP)}_Q}{\Gamma^{(q)}_Q} 
= \frac{1.4\Gamma^{(\mathrm{sat,d})}}{1.1\times 8N_q \Gamma^{(\mathrm{sat,d})}}
= 8.8 \times 10^{-3} \left(\frac{N_q}{18}\right)^{-1}.
\end{equation}

\section{Baryon, axino and MLSP abundances from $Q$-ball decay}
In this section, we estimate the number densities of the baryon and the axino dark matter. 
We also calculate the MLSP abundance. It is constrained by the fact that the produced MLSPs
do not destroy light elements created at BBN. The analysis largely follows Refs.~\cite{KK2011,KKY2013}. 

\subsection{Baryon and axino densities}
The number densities of the baryon, the axino and the MLSP are expressed 
in terms of the AD field number density $n_\phi$ as
\begin{eqnarray}
& & n_{b} \simeq \epsilon b n_{\phi}, \\
\label{na}
& & n_{\tilde{a}} \simeq  B_{\tilde{a}} n_{\phi}, \\
\label{nMLSP}
& & n_{\mathrm{MLSP}} \simeq B_{\rm{MLSP}}\frac{Q_{\rm cr}}{Q} n_{\phi},
\end{eqnarray}
respectively. The ratio of dark matter to baryon energy densities is $\rho_{\textrm{DM}}/\rho_{b} \simeq 5$ 
\cite{Planck2015}, so 
\begin{equation}
\label{dmbaryonratio}
\frac{\rho_{\tilde{a}}}{\rho_{b}} \simeq \frac{m_{\tilde{a}}B_{\tilde{a}}}{m_{N}\epsilon b} \simeq 5.
\end{equation}
This gives an expression for $\epsilon$ such that
\begin{eqnarray}
\label{epsilon}
\epsilon \simeq \frac{m_{\tilde{a}}}{m_{N}}\frac{B_{\tilde{a}}}{5b} 
\simeq \left\{
\begin{array}{ll}
\displaystyle{ 1.8\times 10^{-5} \ b^{-1} \left(\frac{m_{\tilde{a}}}{10\,\rm MeV}\right)\left(\frac{N_q}{18}\right)^{-1} }
 \hspace{22mm}
 \mathrm{(for\ the\ saturated\ case),}\\[5mm]
\displaystyle{1.0\times 10^{-9} \ b^{-1}  \left(\frac{m_{\tilde{a}}}{10\,\rm MeV}\right)  
 \left( \frac{f_a}{10^{12}\,\mathrm{GeV}}\right)^{-2}
\left( \log{\frac{f_{a}}{10^3\, \mathrm{GeV}}} \right)^2 } \\
\hspace{20mm}\displaystyle{\times \left(\frac{M_F}{10^6\,\rm{GeV}}\right)^{2}
\left(\frac{Q}{10^{23}}\right)^{1/2}\left(\frac{\zeta}{2.5}\right)^2}
\\[3mm]\hspace{75mm}
\mathrm{(for\ the\ unsaturated\ case).}
\end{array}
\right.
\end{eqnarray}
Therefore, the orbit of the AD field is typically oblate, and we generally set 
$\beta=6\times 10^{-5}$ below. The baryon number abundance for the non-$Q$-ball 
dominated case (NQD) and the $Q$-ball dominated case (QD) are estimated as \cite{KK2011}
\begin{eqnarray}
\label{Yb}
Y_{b} \equiv \frac{n_{b}}{s} = 
\left\{
\begin{array}{ll}
\displaystyle{\left.\frac{3T_{\mathrm{D}}}{4}\frac{n_{b}}{\rho_{Q}}\right|_{\rm D}
\simeq \frac{3T_{\rm D}}{4} \left. \frac{n_b}{\rho_Q}\right|_{\rm osc} 
\simeq\frac{9T_{\mathrm{D}}\epsilon b}{16\omega_{Q}}}
& \mathrm{(QD),}\\[5mm]
\displaystyle{\left.\frac{3T_{\mathrm{RH}}}{4}\frac{n_b}{\rho_{\mathrm{rad}}}\right|_{\mathrm{RH}}
\simeq \frac{3T_{\rm RH}}{4} \left. \frac{n_b}{\rho_{\rm inf}}\right|_{\rm osc} 
\simeq\frac{9}{8\sqrt{2}}\epsilon b \beta^{-3/4}\frac{M_F T_{\rm RH}}{M_{\rm P}^2} Q^{3/4}}
& \mathrm{(NQD).}
\end{array}
\right.
\end{eqnarray}
We can thus obtain the baryon abundance $Y_b$ as
\begin{eqnarray}
\label{Ybsat}
\left. \frac{Y_b}{10^{-10}} \right|_{\mathrm{sat}}^{\rm NQD}
&\simeq& 6.4 \times 10  \left(\frac{m_{\tilde{a}}}{10\,\rm MeV}\right)   
\left(\frac{M_F}{10^{6}\, \mathrm{GeV}}\right)  
\left(\frac{Q}{10^{23}}\right)^{3/4}\left(\frac{T_{\rm RH}}{10^{7}\,{\rm GeV}}\right) 
\left(\frac{N_q}{18}\right)^{-1}\left(\frac{\beta}{6\times 10^{-5}}\right)^{-3/4},
 \nonumber\\ & & \\
\label{Ybunsat}
\left. \frac{Y_b}{10^{-10}} \right|_{\mathrm{unsat}}^{\rm NQD}
&\simeq& 3.7 \times 10^{-3} \left(\frac{m_{\tilde{a}}}{10\,\rm MeV}\right) 
\left( \frac{f_a}{10^{12}\, \mathrm{GeV}}\right)^{-2} 
\left( \log{\frac{f_a}{10^3 \,\mathrm{GeV}}} \right)^2
\nonumber \\ & & \hspace{10mm} \times  
\left(\frac{M_F}{10^6\,\mathrm{GeV}}\right)^3 \left(\frac{Q}{10^{23}}\right)^{5/4}
\left(\frac{T_{\rm RH}}{10^7\,\mathrm{GeV}}\right) \left(\frac{\beta}{6\times 10^{-5}}\right)^{-3/4}
\left(\frac{\zeta}{2.5}\right)^2.
\end{eqnarray}
respectively for saturated and unsaturated cases in NQD, and 
\begin{eqnarray}
\label{YbsatQD}
\left. \frac{Y_b}{10^{-10}} \right|_{\mathrm{sat}}^{\rm QD}
& \simeq & 1.5 \times 10  \left(\frac{m_{\tilde{a}}}{10\,\rm MeV}\right)   
\left(\frac{M_F}{10^{6}\, \mathrm{GeV}}\right)^{-1} 
\left(\frac{Q}{10^{23}}\right)^{1/4}\left(\frac{T_{\rm D}}{3\,{\rm MeV}}\right) 
\left(\frac{N_q}{18}\right)^{-1}\left(\frac{\zeta}{2.5}\right)^{-1}, \\
\label{YbunsatQD}
\left. \frac{Y_b}{10^{-10}} \right|_{\mathrm{unsat}}^{\rm QD}
&\simeq& 8.7 \times 10^{-4} \left(\frac{m_{\tilde{a}}}{10\,\rm MeV}\right) 
\left( \frac{f_a}{10^{12}\, \mathrm{GeV}}\right)^{-2} \left( \log{\frac{f_a}{10^3 \,\mathrm{GeV}}} \right)^2
\nonumber \\ & & \hspace{20mm} \times  
\left(\frac{M_F}{10^6\,\mathrm{GeV}}\right) \left(\frac{Q}{10^{23}}\right)^{3/4}
\left(\frac{T_{\rm D}}{3\,\mathrm{MeV}}\right)\left(\frac{\zeta}{2.5}\right),
\end{eqnarray}
respectively for saturated and unsaturated cases in QD. 

Notice that the ratio of the energy densities of 
the $Q$ ball and the radiation produced by reheating after inflation
\begin{eqnarray}
\label{QD/NQD}
\left.\frac{\rho_{Q}}{\rho_{\mathrm{rad}}} \right|_{\rm D} 
& \simeq & \left.\frac{\rho_{Q}}{\rho_{\mathrm{rad}}}\right|_{\mathrm{RH}}\frac{T_{\mathrm{RH}}}{T_{\rm D}} 
\simeq \frac{Y^{\mathrm{NQD}}_{b}}{Y^{\mathrm{QD}}_{b}}, \nonumber \\
&=& 0.188 \left( \frac{M_F}{10^6\,\mathrm{GeV}}\right)^{3/2}\left( \frac{Q}{10^{23}}\right)^{9/8}\left( \frac{T_{\mathrm{RH}}}{10^7\,\mathrm{GeV}}\right) 
\nonumber\\ & & \times
\left( \frac{N_q}{18}\right)^{-1/2}\left( \frac{N_d}{10.75}\right)^{1/4}\left( \frac{\beta}{6\times10^{-5}}\right)^{-3/4}\left( \frac{\zeta}{2.5}\right)^{1/2},
\end{eqnarray}
determines if the $Q$ balls dominate the universe at the decay time. 
Here, we used Eq.(\ref{Yb}).

\subsection{MLSP density}
Now let us calculate the abundance of MLSPs which are produced by the $Q$-ball decay. We can
estimate it as
\begin{eqnarray}
\label{rhomfq}
\frac{\rho_{\mathrm{MLSP}}}{s} &=& 
m_{\tilde{a}} Y_{\tilde{a}} \frac{\rho_{\mathrm{MLSP}}}{\rho_{\tilde{a}}} 
\nonumber \\ &\simeq& 
5 m_N Y_b \frac{m_{\mathrm{MLSP}} n_{\mathrm{MLSP}}}{m_{\tilde{a}} n_{\tilde{a}}} 
\nonumber \\ &\simeq& 
5 m_N Y_b  \frac{m_{\mathrm{MLSP}}}{m_{\tilde{a}}}4\pi^4
\left( \frac{M_{F}}{m_{\mathrm{MLSP}}}\right)^4 \zeta^4 \frac{1}{Q}  \frac{B_{\rm{MLSP}}}{B_{\tilde{a}}},
\end{eqnarray}
where Eqs.(\ref{na}) and (\ref{nMLSP}) are used in the last line. 
In the saturated case, this becomes,
\begin{eqnarray}
\label{rhomfqsat}
\left. \frac{\rho_{\mathrm{MLSP}}}{s} \right|_{\mathrm{sat}} 
&\simeq& 
2.8\times 10^{-10}\,\mathrm{GeV}
\left(\frac{Y_{b}}{10^{-10}} \right) 
\left( \frac{m_{\tilde{a}}}{10\,\rm MeV}\right)^{-1} 
\left( \frac{m_{\mathrm{MLSP}}}{300\, \mathrm{GeV}} \right)^{-3} 
\nonumber\\ & & \hspace{15mm} 
\times\left(\frac{M_{F}}{10^6\,\mathrm{GeV}}\right)^4 
\left(\frac{Q}{10^{23}}\right)^{-1}\left(\frac{\zeta}{2.5}\right)^4,
\end{eqnarray}
where Eqs.(\ref{branchsat}) and (\ref{branchMLSP}) are used, and, in the unsaturated case, we have,
using Eqs.(\ref{branchunsat}) and (\ref{branchMLSP}), 
\begin{eqnarray}
\label{rhomfqunsat}
\left.\frac{\rho_{\mathrm{MLSP}}}{s}\right|_{\mathrm{unsat}} 
&\simeq& 
4.9 \times10^{-6} \,\mathrm{GeV} 
\left(\frac{Y_{b}}{10^{-10}} \right) 
\left( \frac{m_{\tilde{a}}}{10\,\rm MeV}\right)^{-1} 
\left( \frac{m_{\mathrm{MLSP}}}{300 \,\mathrm{GeV}} \right)^{-3} \left(\frac{N_q}{18}\right)^{-1}
\nonumber\\ & & \hspace{1mm} \times
\left( \frac{f_a}{10^{12}\,\mathrm{GeV}}\right)^2
\left(\log{\left(\frac{f_{a}}{10^3 \,\mathrm{GeV}}\right)}\right)^{-2} 
\left(\frac{M_{F}}{10^6\,\mathrm{GeV}}\right)^2
\left(\frac{Q}{10^{23}}\right)^{-3/2}\left(\frac{\zeta}{2.5}\right)^2.
\end{eqnarray}
If the MLSP abundance from $Q$-ball decay is large, the annihilation takes place and 
the abundance settles down to the annihilation abundances of MLSPs \cite{DM}. 
Those abundances are given by \cite{KKY2013}
\begin{equation}
\label{annihilationbino}
\left. \frac{\rho^{(\rm{ann})}_{\mathrm{MLSP}}}{s}\right|_{\bar{B}} \simeq 6.7\times10^{-5}\,\mathrm{GeV} 
\left(\frac{m_{\rm{MLSP}}}{300\,\rm{GeV}}\right)^{3}\left(\frac{T_{\rm D}}{3\,\rm{MeV}}\right)^{-1}
\left(\frac{N_d}{10.75}\right)^{-1/2},
\end{equation}
\begin{equation}
\label{annihilationstau}
\left. \frac{\rho^{(\rm{ann})}_{\mathrm{MLSP}}}{s}\right|_{\bar{\tau}} \simeq 1.3\times10^{-6}\,\mathrm{GeV} 
\left(\frac{m_{\rm{MLSP}}}{300\,\rm{GeV}}\right)^{3}\left(\frac{T_{\rm D}}{3\,\rm{MeV}}\right)^{-1}
\left(\frac{N_d}{10.75}\right)^{-1/2},
\end{equation}
for bino and stau MLSPs, respectively.

In addition, MLSPs may be produced thermally in the primordial universe. 
We adopt the amount of the primordial bino and stau MLSPs in \cite{BBN}, respectively as
\begin{equation}
\label{rhopribino}
\left. \frac{\rho_{\rm MLSP}}{s}\right|_{\rm pri}^{\tilde{B}} 
= 8 \times 10^{-10}\,\mathrm{GeV} \left( \frac{m_{\rm MLSP}}{300\,\rm GeV}\right)^2,
\end{equation}
\begin{equation}
\label{rhopristau}
\left. \frac{\rho_{\rm MLSP}}{s}\right|_{\rm pri}^{\tilde{\tau}} 
= 6 \times 10^{-11}\,\mathrm{GeV} \left( \frac{m_{\rm MLSP}}{300\,\rm GeV}\right)^2.
\end{equation}

The upper limit on the MLSP abundance is given by the fact that the decay of the MLSPs  
should not affect abundances of light elements synthesized during the BBN. We assume that the MLSP
is the bino or the stau. The upper bound can then be estimated approximately as \cite{BBN}
\begin{eqnarray}
\label{BBNconbino}
\left.\frac{\rho_{\mathrm{MLSP}}}{s}\right|_{\rm bino}
&\lesssim&\left\{
\begin{array}{ll}
\displaystyle{
5 \times 10^{-9}\,\mathrm{GeV}}
& \left( 0.1 \sec \lesssim \tau_{\mathrm{MLSP}} \lesssim 80 \sec \right), \\[2mm]
\displaystyle{
1 \times 10^{-13}\,\mathrm{GeV}}\ 
& \left( 80 \sec \lesssim \tau_{\mathrm{MLSP}}  \right),
\end{array}
\right.\\[2mm]
\label{BBNconstau}
\left.\frac{\rho_{\mathrm{MLSP}}}{s}\right|_{\rm stau}
&\lesssim&\left\{
\begin{array}{ll}
\displaystyle{
5 \times 10^{-6}\,\mathrm{GeV}}
& \left( 2 \sec \lesssim \tau_{\mathrm{MLSP}} \lesssim 60 \sec \right), \\[2mm]
\displaystyle{
6 \times 10^{-10}\,\mathrm{GeV}}\ 
& \left( 60 \sec \lesssim \tau_{\mathrm{MLSP}} \lesssim 4 \times 10^3 \sec \right), \\[2mm]
1 \times 10^{-13}\,\mathrm{GeV}
& \left( 4 \times 10^3 \sec \lesssim \tau_{\mathrm{MLSP}} \right),
\end{array}
\right.
\end{eqnarray}
for the bino and stau MLSP cases, respectively.
Here $\tau_{\mathrm{MLSP}}$ is the life time of the MLSP and 
$m_{\mathrm{MLSP}} = 300\,\mathrm{GeV} $ is assumed. 

From the decay rate of the bino into the axino [Eq.(\ref{GMLSP})], 
and gravitino [Eq.(\ref{binogravitino})], the lifetimes are approximately calculated respectively as
\begin{eqnarray}
\label{lifetimebinoaxi}
& & \tau_{\mathrm{MLSP}\rightarrow \tilde{a}}^{\tilde{B}} = 1.1\,\sec 
\left( \frac{m_{\rm MLSP}}{300\,\mathrm{GeV}}\right)^{-3} 
\left( 1-\left( \frac{m_{\tilde{a}}}{m_{\mathrm{MLSP}}}\right)^2\right)^{-3}
\left(\frac{f_{a}}{10^{12}\,\mathrm{GeV}}\right)^2, \\
\label{lifetimebinogra}
& & 
\tau_{\mathrm{MLSP}\rightarrow 3/2}^{\tilde{B}} = 
\displaystyle{3.1\times10^{-2} \,\sec \left( \frac{m_{3/2}}{10\,\mathrm{MeV}}\right)^2 
\left(\frac{m_{\mathrm{MLSP}}}{300\,\mathrm{GeV}}\right)^{-5} }  \nonumber\\
& & \hspace{40mm} \times
\displaystyle{\left( 1-\left( \frac{m_{3/2}}{m_{\mathrm{MLSP}}}\right)^2\right)^{-3} \left( 1+3\left( \frac{m_{3/2}}{m_{\mathrm{MLSP}}}\right)^2\right)^{-1}},
\end{eqnarray}
and, from Eqs.(\ref{stauaxino}) and (\ref{staugravitino}), 
the lifetimes of the stau MLSP decay into axino and gravitino are given as
\begin{eqnarray}
\label{lifetimestauaxi}
& & \tau_{\mathrm{MLSP}\rightarrow \tilde{a}}^{\tilde{\tau}} = 4.7 \times 10^3\,\sec
\left( \frac{m_{\rm MLSP}}{300\,\mathrm{GeV}}\right)^{-1}
\left( 1-\left( \frac{m_{\tilde{a}}}{m_{\mathrm{MLSP}}}\right)^2\right)^{-2} \nonumber \\
& & \hspace{50mm} \times \left(\frac{f_{a}}{10^{12}\,\mathrm{GeV}}\right)^2
\left( \log{\left(\frac{f_{a}}{\sqrt{2} m_{\rm MLSP}}\right)}\right)^{-2},\\
\label{lifetimestaugra}
& & \tau_{\mathrm{MLSP}\rightarrow 3/2}^{\tilde{\tau}} = 2.3\times 10^{-2} \,\sec 
\left( \frac{m_{3/2}}{10\,\mathrm{MeV}}\right)^2 
\left(\frac{m_{\mathrm{MLSP}}}{300\,\mathrm{GeV}}\right)^{-5} 
\left( 1-\left( \frac{m_{3/2}}{m_{\mathrm{MLSP}}}\right)^2\right)^{-4},
\end{eqnarray}
respectively. Here and hereafter, we assume $m_{\tilde{a}} \simeq m_{3/2}$ \footnote{
Although $m_{\tilde{a}} \simeq m_{3/2}$ is natural, 
the axino mass may vary large depending on the actual models \cite{axinomass}. 
When $m_{\tilde{a}} < m_{3/2}$, the axino is the LSP to be dark matter, while the 
$Q$-ball decay into gravitinos can be neglected because $\frac{B_{\tilde{a}}}{B_{3/2}} \gg 1$ 
[Eq.(\ref{branchingaxinotogravitino})]. On the other hand, when $m_{\tilde{a}} > m_{3/2}$, 
even though the gravitino is the LSP, the axino still plays a role as dark matter, if the life time 
of the axino decay into gravitino and axion is longer than the age of the universe, 
$\tau_{\tilde{a}} > t_{0}$. In the $\tau_{\tilde{a}} < t_{0}$ case, relativistic axions produced 
by the decay may effect the evolution of the universe.}
. We then obtain the upper bound of 
the MLSP abundance by using the smaller lifetime of $\tau_{\mathrm{MLSP}\rightarrow \tilde{a}}$ and 
$\tau_{\mathrm{MLSP}\rightarrow 3/2}.$

\begin{figure}
\begin{center}
\begin{tabular}{cc}
\includegraphics[width=85mm]{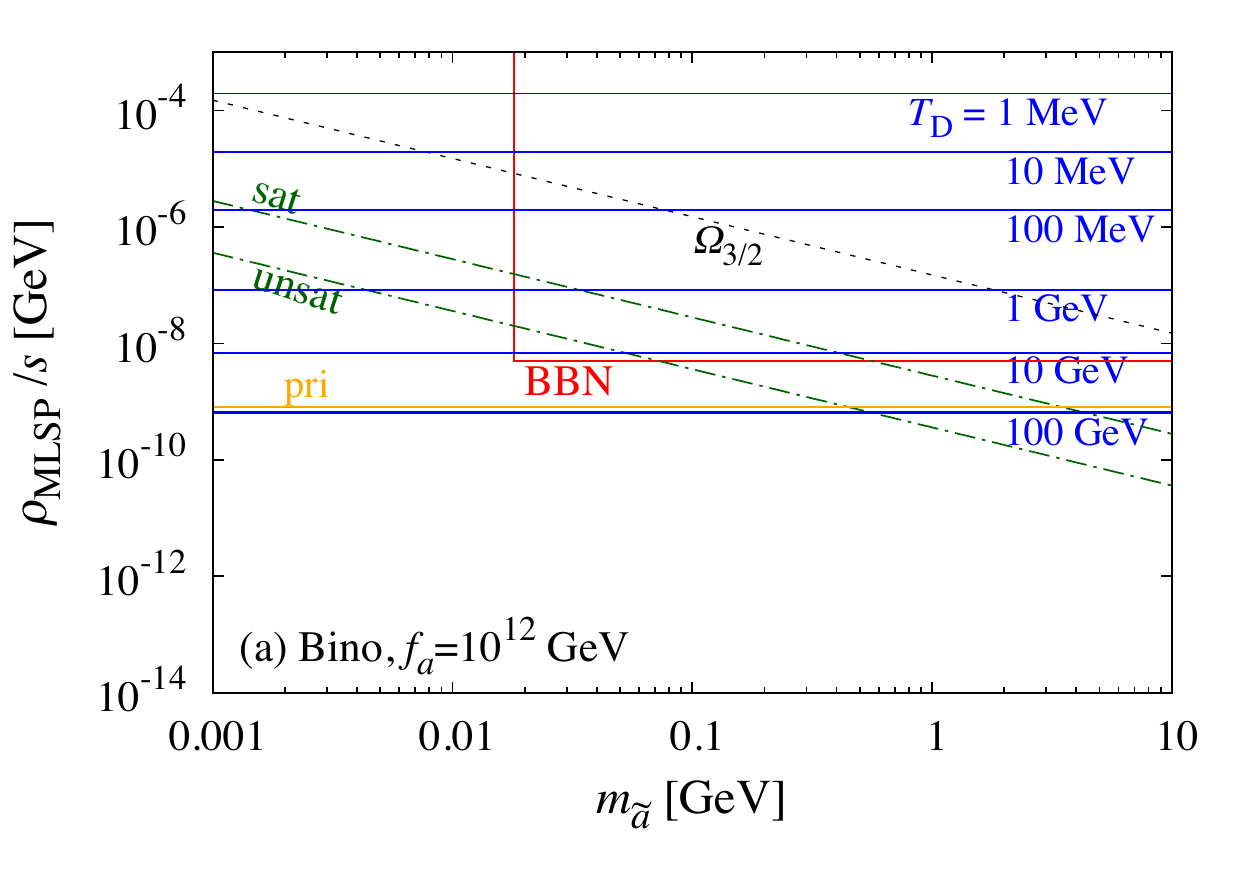} &
\includegraphics[width=85mm]{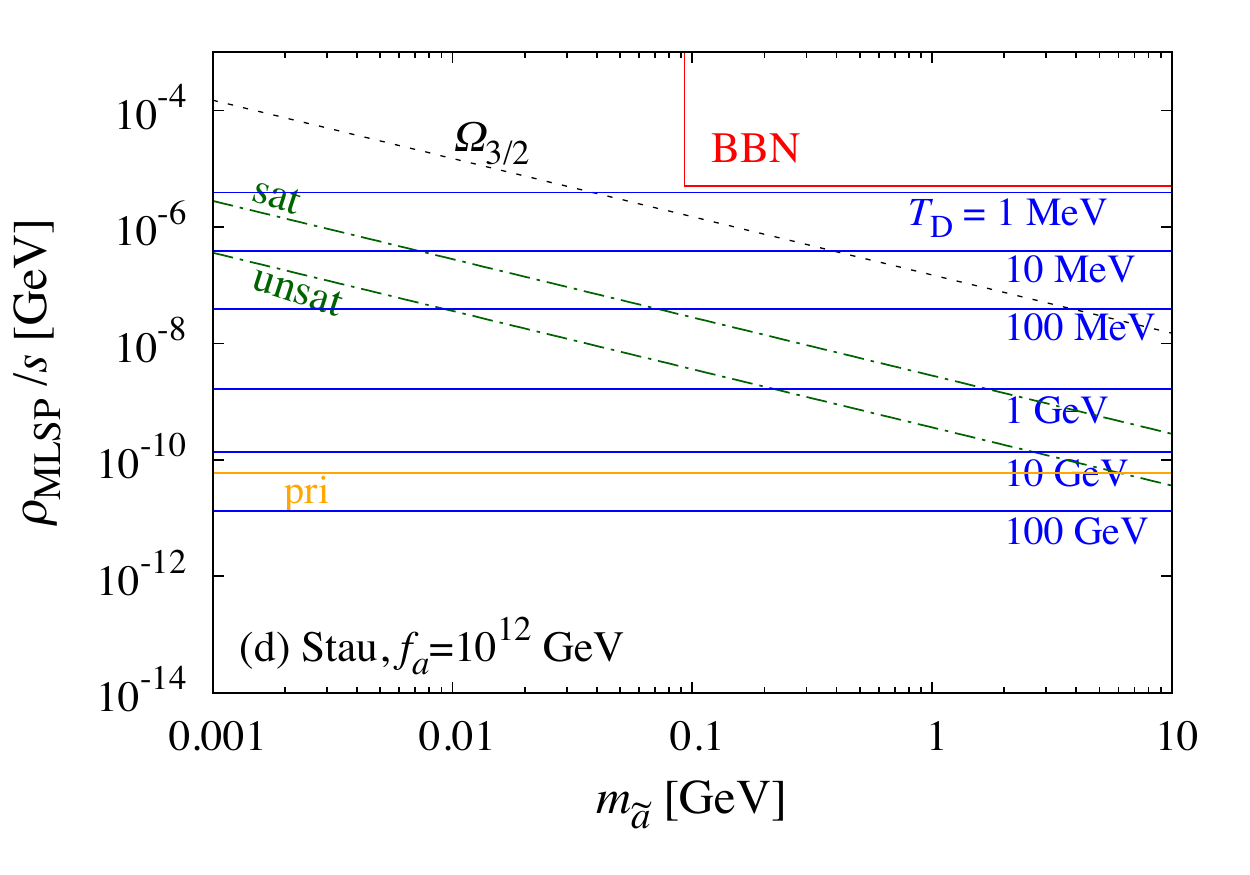} \\
\includegraphics[width=85mm]{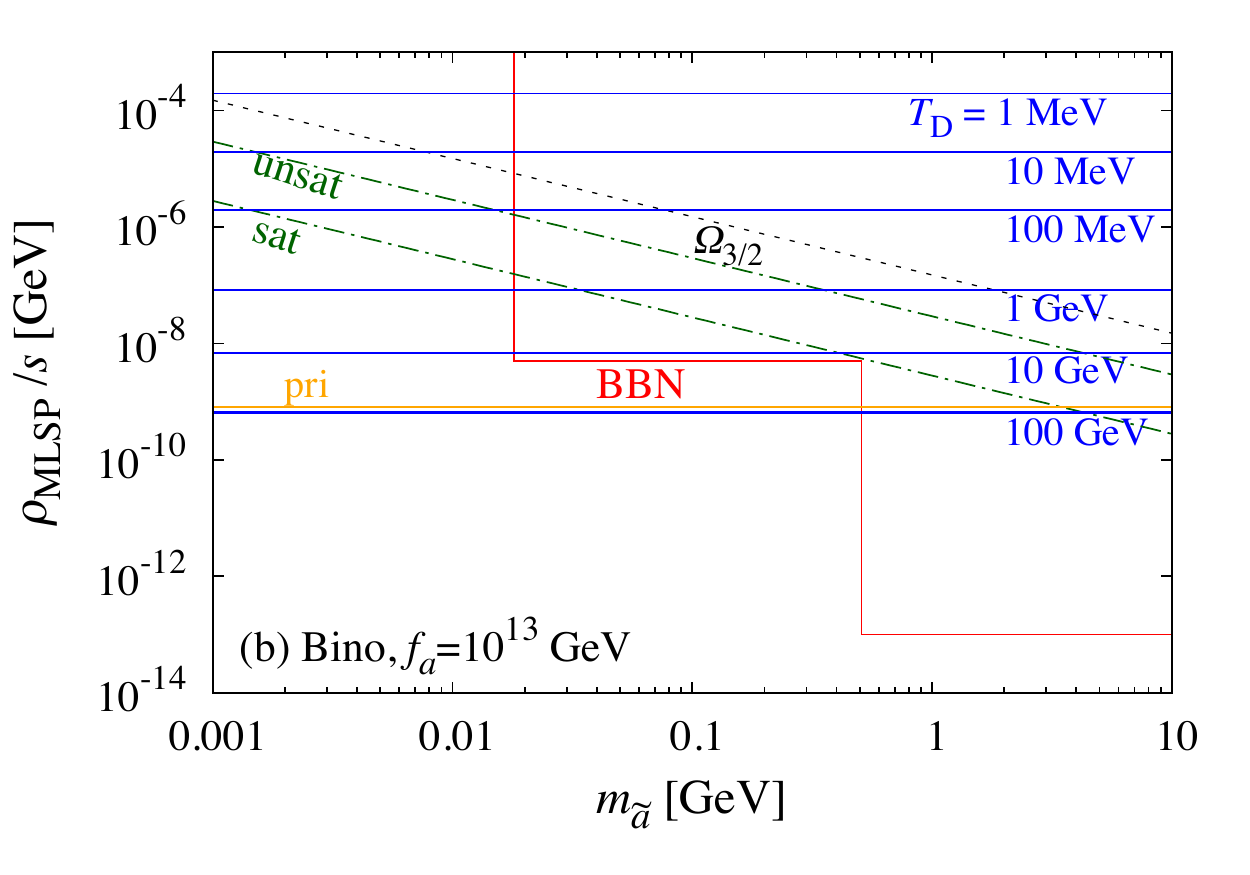} &
\includegraphics[width=85mm]{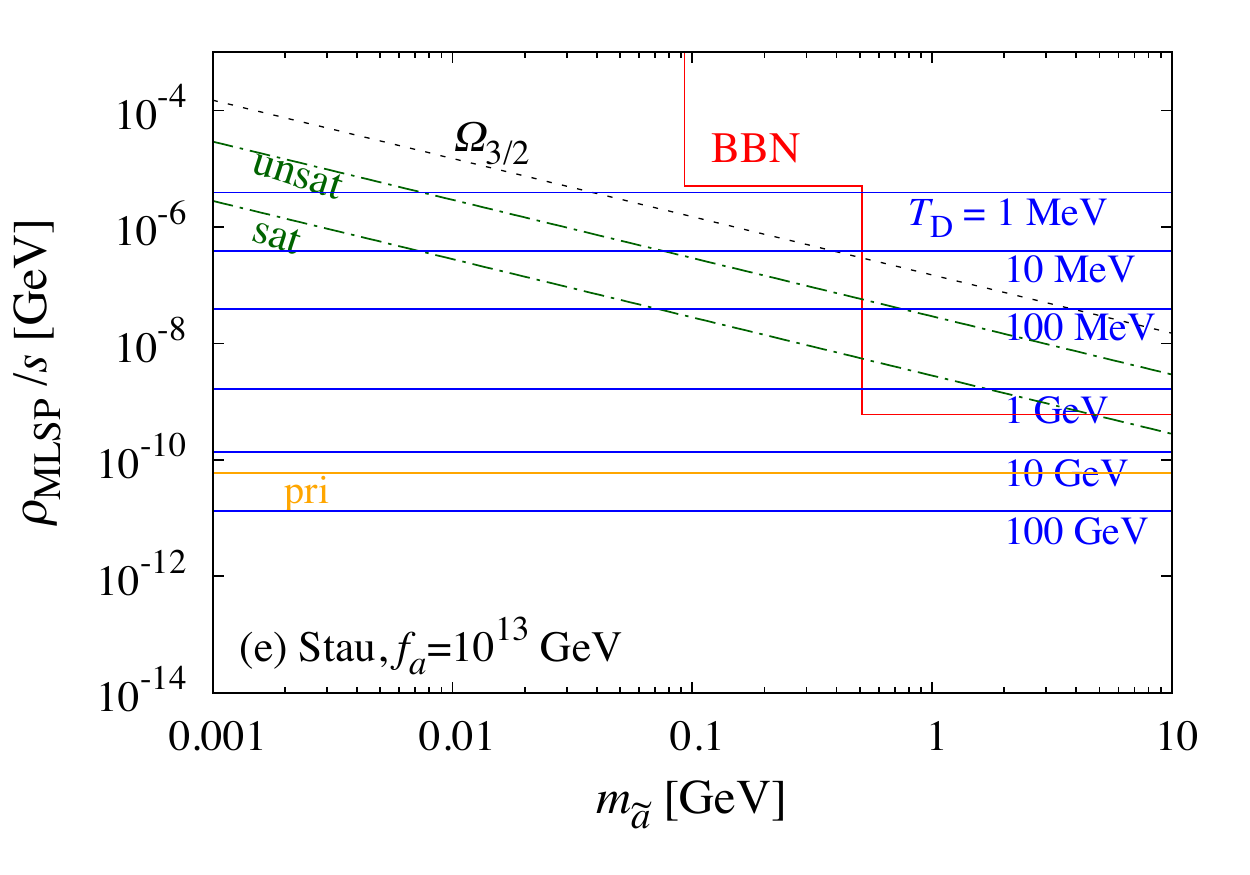} \\
\includegraphics[width=85mm]{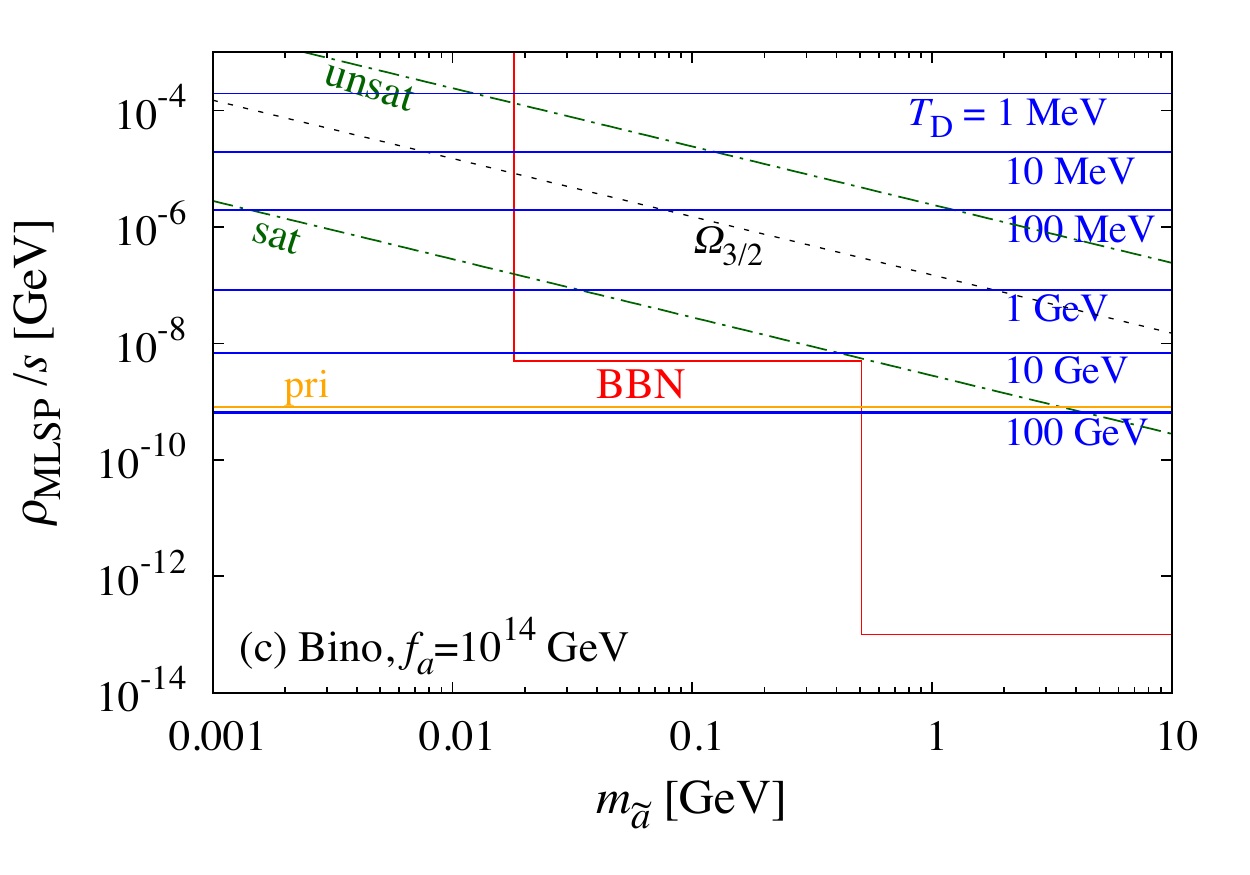} &
\includegraphics[width=85mm]{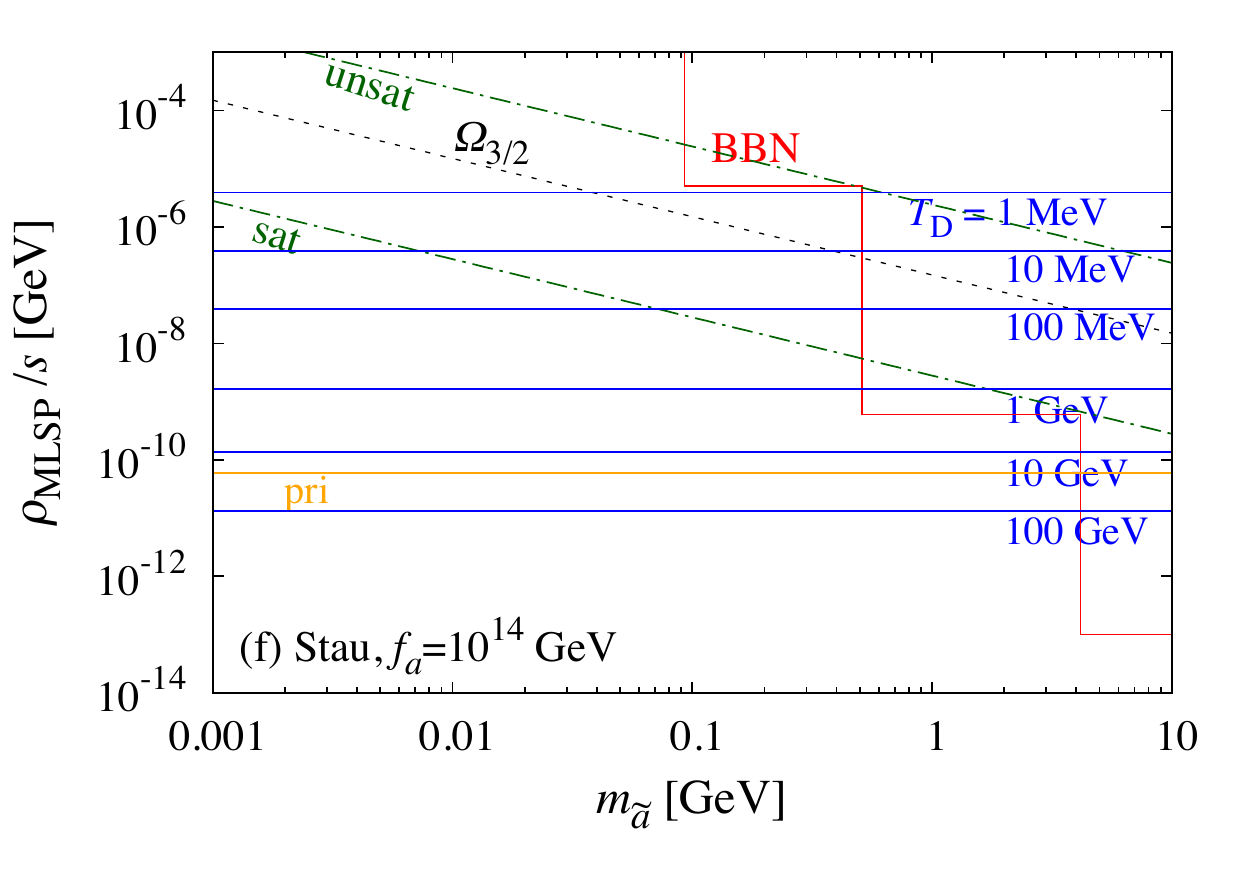}
\end{tabular}
\caption{BBN constraints on the total MLSP abundance for the bino and stau MLSP 
($m_{\rm MLSP}=300$~GeV) for $f_{a}=10^{12},10^{12}, 10^{13}$\,GeV. We set 
$M_F=10^7$~GeV and $Q=10^{24}$ as the typical values.}
\label{rhoMLSP}
\end{center}
\end{figure}

We plot the MLSP abundance in Fig.~\ref{rhoMLSP}. Green lines denote the abundance from 
the $Q$-ball decay (\ref{rhomfqsat}) and (\ref{rhomfqunsat}) for $M_F=10^7$~GeV and $Q=10^{24}$.
Blue lines show the annihilation density for the bino (\ref{annihilationbino}) or stau (\ref{annihilationstau})
MLSPs for some different $T_{\rm D}$. Orange lines represent the primordial abundance 
for the bino (\ref{rhopribino}) or the stau (\ref{rhopristau}). Red lines are the upper limits (\ref{BBNconbino})
or (\ref{BBNconstau}). Notice that there is no BBN limit for $f_a\lesssim10^{11}$~GeV in both the bino and stau
MLSP cases. Black dotted line is the abundance that the MLSP decay gives the right amount of the gravitino 
dark matter. We can see that typically $m_{\tilde{a}} \lesssim 10$~MeV is allowed for the 
bino MLSP case. On the other hand, for the stau MLSP case, it is allowed for $m_{\tilde{a}} \lesssim$~GeV 
for larger $f_a$, while there is no limit for $f_a\lesssim10^{12}$~GeV. In any case, the allowed range of 
$m_{\tilde{a}}$ becomes larger if the decay temperature $T_{\rm D}$ is high enough.

\section{Constraints on model parameters}
In this section, we investigate the allowed region for the $Q$-ball parameters ($Q$, $M_F$). 
Our scenario must explain the amounts of both the baryon asymmetry and the dark matter from the 
$Q$-ball decay. We thus obtain the charge $Q$ in terms of $M_F$, from Eqs.(\ref{YbsatQD}) or 
(\ref{YbunsatQD}) with Eq.(\ref{td}), as
\begin{eqnarray}
\label{Qsat2QD}
& &
Q_{\mathrm{sat}}^{\mathrm{QD}}
\simeq 5.6 \times 10^{29}
\left(\frac{Y_b}{10^{-10}}\right)^{-8/3}
\left( \frac{m_{\tilde{a}}}{10\,\rm MeV}\right)^{8/3}
\left( \frac{M_F}{10^6\, \mathrm{GeV}}\right)^{-4/3}
\left( \frac{N_q}{18}\right)^{-4/3}
\nonumber \\ & & \hspace{95mm} \times
\left(\frac{N_d}{10.75}\right)^{-2/3}
\left( \frac{\zeta}{2.5}\right)^{-4/3}, \\
\label{Qunsat2QD}
& &
Q_{\mathrm{unsat}}^{\mathrm{QD}}
\simeq 4.9 \times 10^{36}
\left(\frac{Y_b}{10^{-10}}\right)^{8} 
\left( \frac{m_{\tilde{a}}}{10\,\rm MeV}\right)^{-8}
\left( \frac{f_a}{10^{12}\,\mathrm{GeV}}\right)^{16} 
\left( \log\left(\frac{f_{a}}{10^3 \,\mathrm{GeV}}\right) \right)^{-16}
 \nonumber \\
& & \hspace{65mm}\times 
\left( \frac{M_F}{10^6 \, \mathrm{GeV}}\right)^{-12}
\left(\frac{N_q}{18}\right)^{-4}
\left(\frac{N_d}{10.75}\right)^{2}
\left( \frac{\zeta}{2.5} \right)^{-12},
\end{eqnarray}
for the QD case. On the other hand, in the NQD case, Eqs.(\ref{Ybsat}) or (\ref{Ybunsat}) give the charge $Q$
in terms of $M_F$ and the reheating temperature $T_{\rm RH}$ as
\begin{eqnarray}
\label{Qsat2}
& & 
Q_{\mathrm{sat}}^{\mathrm{NQD}}
\simeq 3.9 \times 10^{20}
\left(\frac{Y_b}{10^{-10}}\right)^{4/3}
\left( \frac{m_{\tilde{a}}}{10\,\rm MeV}\right)^{-4/3}
\left( \frac{M_F}{10^6\, \mathrm{GeV}}\right)^{-4/3}
\left( \frac{T_{\rm RH}}{10^7 \,\mathrm{GeV}}\right)^{-4/3}
\nonumber \\ & & \hspace{95mm} \times 
\left( \frac{N_q}{18}\right)^{4/3}
\left( \frac{\beta}{6\times 10^{-5}}\right), \\
\label{Qunsat2}
& & 
Q_{\mathrm{unsat}}^{\mathrm{NQD}}
\simeq 8.9 \times 10^{24} 
\left(\frac{Y_b}{10^{-10}}\right)^{4/5}
\left( \frac{m_{\tilde{a}}}{10\,\rm MeV}\right)^{-4/5}
\left( \frac{f_a}{10^{12}\,\mathrm{GeV}}\right)^{8/5} 
\left( \log\left(\frac{f_{a}}{10^3 \,\mathrm{GeV}}\right) \right)^{-8/5}
\nonumber \\ & & \hspace{35mm}\times 
\left( \frac{M_F}{10^6 \, \mathrm{GeV}}\right)^{-12/5}
\left( \frac{T_{\rm RH}}{10^7\, \mathrm{GeV}}\right)^{-4/5}
\left( \frac{\beta}{6\times 10^{-5}}\right)^{3/5}
\left( \frac{\zeta}{2.5} \right)^{-8/5}.
\end{eqnarray}
In this case, the largest reheating temperature gives the lower limit on $Q$, while the smallest temperature
leads to the upper bound. Since the smallest possible reheating temperature is simply given by 
$T_{\rm RH, min} = T_{\rm D}$, we obtain the upper bound as
\begin{eqnarray}
\label{Qsat2NQDU}
& & 
Q_{\rm sat, U}^{\rm NQD} \simeq 8.7 \times 10^{73} 
\left(\frac{Y_b}{10^{-10}}\right)^8
\left( \frac{m_{\tilde{a}}}{10\,\rm MeV}\right)^{-8}
\left( \frac{M_F}{10^6\, \mathrm{GeV}}\right)^{-12}
\nonumber \\ & & \hspace{50mm} \times 
\left( \frac{\beta}{6\times 10^{-5}}\right)^6
\left( \frac{N_q}{18}\right)^4
\left( \frac{N_d}{10.75}\right)^2
\left( \frac{\zeta}{2.5}\right)^{-4},\\
\label{Qunsat2NQDU}
& & 
Q_{\rm unsat, U}^{\rm NQD} \simeq 9.5 \times 10^{39}
\left(\frac{Y_b}{10^{-10}}\right)^{8/5}
\left( \frac{m_{\tilde{a}}}{10\,\rm MeV}\right)^{-8/5}
\left( \frac{f_a}{10^{12}\,\mathrm{GeV}}\right)^{16/5} 
\left( \log\left(\frac{f_{a}}{10^3 \,\mathrm{GeV}}\right) \right)^{-16/5}
\nonumber \\ & & \hspace{30mm}\times 
\left( \frac{M_F}{10^6 \, \mathrm{GeV}}\right)^{-28/5}
\left( \frac{\beta}{6\times 10^{-5}}\right)^{6/5}
\left( \frac{N_q}{18}\right)^{-4/5}
\left( \frac{N_d}{10.75}\right)^{2/5}
\left( \frac{\zeta}{2.5} \right)^{-4}.
\end{eqnarray}

The largest possible reheating temperature is obtained from the fact that 
the dark matter consists of the axinos produced by the $Q$-ball decay and the thermally produced
gravitinos or axinos cannot be the dominant component of the dark matter:
\begin{equation}
\label{TH}
\max(\Omega_{\tilde{a}}^{\mathrm{TH}}h^2, \Omega_{3/2}^{\mathrm{TH}}h^2)
 \lesssim \Omega_{\rm DM}h^2 \simeq 0.11,
\end{equation}
where $\Omega_{\tilde{a}}^{\mathrm{TH}}$ and $\Omega_{3/2}^{\mathrm{TH}}$ 
respectively denote the density parameters of thermally produced axino and gravitino,  
and $h$ is the Hubble constant in units of $100\ \mathrm{km/s/Mpc}$. 
The constraint from the gravitino is written as \cite{Kawasaki:2006hm},
\begin{eqnarray}
\label{tmaxg}
T_{\rm RH} \lesssim T_{\rm RH,max}^{(3/2)} 
&\simeq& 7.5 \times 10^4\,\mathrm{GeV} 
\left( \frac{m_{\tilde{a}}}{10\,\rm MeV}\right)\left( \frac{m_{\tilde{g}}}{1\,\mathrm{TeV}}\right)^{-2},
\end{eqnarray}
where we use $m_{3/2} \simeq m_{\tilde{a}}$ as before.
Meanwhile, the constraint from the axino depends on axion models. 
Here we consider two classes of axion models: the KSVZ \cite {KSVZ} and the DFSZ \cite{DFSZ} models. 
In the KSVZ model, the thermally produced axino density parameter is estimated as 
Eq.(\ref{appOmegaKSVZ}), so that the constraint (\ref{TH}) leads to
\begin{equation}
\label{tmaxk}
T_{\rm RH}  \lesssim T_{\rm RH,max}^{({\rm KSVZ} \ \tilde{a})} \equiv 
1.1 \times 10^6\ \mathrm{GeV}\left(\frac{m_{\tilde{a}}}{10\,\rm MeV}\right)^{-1}
\left( \frac{f_{a}}{10^{12}\,\mathrm{GeV}}\right)^2.
\end{equation}
In the DFSZ model, the higgsino decay through the axino-Higgsino-Higgs interaction is dominant 
[Eq.(\ref{appOmegaDFSZ1})] at the low reheating temperature 
$(T_{\rm RH} \lesssim 5\times10^7 \,\mathrm{GeV})$ \cite{Chun}, while the axino production 
by scatterings is dominant [Eq.(\ref{appOmegaDFSZ2})] at the high reheating temperature 
$(T_{\rm RH} \gtrsim 5\times 10^7\,\mathrm{GeV})$. Thus the abundance of thermally 
produced axinos can be estimated as \cite{Choi:2011yf}
\begin{equation}
\label{thermalaxino}
\Omega_{\tilde{a}}^{\mathrm{TH}}h^{2} \simeq 
\left\{
\begin{array}{ll}
\displaystyle{
0.5 \left( \frac{m_{\tilde{a}}}{10\,\rm MeV}\right) 
\left( \frac{f_{a}}{10^{12}\,\mathrm{GeV}}\right)^{-2}} &  
\left( T_{\rm RH} \lesssim 5\times10^7\, \mathrm{GeV}\right), \\[3mm] 
\displaystyle{
0.1 \left(\frac{m_{\tilde{a}}}{10\,\rm MeV}\right) 
\left( \frac{f_{a}}{10^{12}\,\mathrm{GeV}}\right)^{-2} 
\left( \frac{T_{\rm RH}}{10^7\,\mathrm{GeV}}\right)} & 
\left( T_{\rm RH} \gtrsim 5\times10^7 \,\mathrm{GeV}\right).
\end{array}
\right.
\end{equation}
For $T_{\mathrm{RH}} \gtrsim 5\times10^7\,\mathrm{GeV}$, we obtain the upper limit on $T_{\rm RH}$ as
\begin{equation}
\label{tmaxv}
T_{\mathrm{RH}} \lesssim T_{\rm RH,max}^{({\rm DFSZ} \ \tilde{a})} 
\equiv 1.1 \times 10^7\,\mathrm{GeV} 
\left( \frac{m_{\tilde{a}}}{10\,\rm MeV}\right)^{-1} 
\left( \frac{f_{a}}{10^{12}\,\mathrm{GeV}}\right)^2,
\end{equation}
while, for $T_{\rm RH} \lesssim 5\times 10^7\,\mathrm{GeV}$, the condition (\ref{TH})
only leads to the constraint on $f_{a}$ and $m_{\tilde{a}}$ as
\begin{equation}
\label{DFSZcon}
\left( \frac{m_{\tilde{a}}}{10\,\rm MeV}\right) 
\left( \frac{f_a}{10^{12}\,\mathrm{GeV}}\right)^{-2} \lesssim 0.22.
\end{equation}
Therefore, the largest possible reheating temperature is obtained as
\begin{equation}
T_{\rm RH,max} = \min\left(T_{\rm RH,max}^{(3/2)}, 
T_{\rm RH,max}^{({\rm KSVZ} \ \tilde{a})} \, {\rm or} \ T_{\rm RH,max}^{({\rm DFSZ} \ \tilde{a})} \right).
\end{equation}
Inserting Eqs.(\ref{tmaxg}), (\ref{tmaxk}) or (\ref{tmaxv}) into Eqs.(\ref{Qsat2}) and (\ref{Qunsat2}),
we get the lower bound as
\begin{eqnarray}
\label{Qsat2NQDL}
& & 
Q_{\rm sat, L}^{\rm NQD(3/2)}
\simeq 2.7 \times 10^{23}
\left(\frac{Y_b}{10^{-10}}\right)^{4/3}
\left( \frac{m_{\tilde{a}}}{10\,\rm MeV}\right)^{-8/3}
\left( \frac{M_F}{10^6\, \mathrm{GeV}}\right)^{-4/3}
\nonumber \\ & & \hspace{50mm} \times 
\left( \frac{N_q}{18}\right)^{4/3}
\left( \frac{\beta}{6\times 10^{-5}}\right)
\left( \frac{m_{\tilde{g}}}{1\,\rm TeV}\right)^{8/3}, \\
& & 
Q_{\rm unsat, L}^{\rm NQD(3/2)}
\simeq 4.5 \times 10^{26} 
\left(\frac{Y_b}{10^{-10}}\right)^{4/5}
\left( \frac{m_{\tilde{a}}}{10\,\rm MeV}\right)^{-8/5}
\left( \frac{f_a}{10^{12}\,\mathrm{GeV}}\right)^{8/5} 
\left( \log\left(\frac{f_{a}}{10^3 \,\mathrm{GeV}}\right) \right)^{-8/5}
\nonumber \\ & & \hspace{35mm}\times 
\left( \frac{M_F}{10^6 \, \mathrm{GeV}}\right)^{-12/5}
\left( \frac{\beta}{6\times 10^{-5}}\right)^{3/5}
\left( \frac{\zeta}{2.5} \right)^{-8/5}
\left( \frac{m_{\tilde{g}}}{1\,\rm TeV}\right)^{8/5},
\end{eqnarray}
\begin{eqnarray}
& & 
Q_{\rm sat, L}^{\rm NQD(KSVZ)}
\simeq 7.4 \times 10^{21}
\left(\frac{Y_b}{10^{-10}}\right)^{4/3}
\left( \frac{f_a}{10^{12}\, \mathrm{GeV}}\right)^{-8/3}
\left( \frac{M_F}{10^6\, \mathrm{GeV}}\right)^{-4/3}
\nonumber \\ & & \hspace{90mm} \times 
\left( \frac{N_q}{18}\right)^{4/3}
\left( \frac{\beta}{6\times 10^{-5}}\right), \\
&&
Q_{\rm unsat, L}^{\rm NQD(KSVZ)}
\simeq 5.2 \times 10^{25}
\left(\frac{Y_b}{10^{-10}}\right)^{2/3}
\left( \log\left(\frac{f_{a}}{10^3 \,\mathrm{GeV}}\right) \right)^{-8/5}
\left( \frac{M_F}{10^6 \, \mathrm{GeV}}\right)^{-12/5}
\nonumber \\ & & \hspace{90mm}\times 
\left( \frac{\beta}{6\times 10^{-5}}\right)^{3/5}
\left( \frac{\zeta}{2.5} \right)^{-8/5}, 
\end{eqnarray}
\begin{eqnarray}
& & 
Q_{\rm sat, L}^{\rm NQD(DFSZ)}
\simeq 3.4 \times 10^{20}
\left(\frac{Y_b}{10^{-10}}\right)^{4/3}
\left( \frac{f_a}{10^{12}\, \mathrm{GeV}}\right)^{-8/3}
\left( \frac{M_F}{10^6\, \mathrm{GeV}}\right)^{-4/3}
\nonumber \\ & & \hspace{90mm} \times 
\left( \frac{N_q}{18}\right)^{4/3}
\left( \frac{\beta}{6\times 10^{-5}}\right), \\
\label{Qunsat2NQDL}
&&
Q_{\rm unsat, L}^{\rm NQD(DFSZ)}
\simeq 8.2 \times 10^{24}
\left(\frac{Y_b}{10^{-10}}\right)^{4/5}
\left( \log\left(\frac{f_{a}}{10^3 \,\mathrm{GeV}}\right) \right)^{-8/5}
\left( \frac{M_F}{10^6 \, \mathrm{GeV}}\right)^{-12/5}
\nonumber \\ & & \hspace{90mm}\times 
\left( \frac{\beta}{6\times 10^{-5}}\right)^{3/5}
\left( \frac{\zeta}{2.5} \right)^{-8/5}.
\end{eqnarray}

\begin{figure}
\begin{center}
\begin{tabular}{cc}
\includegraphics[width=85mm]{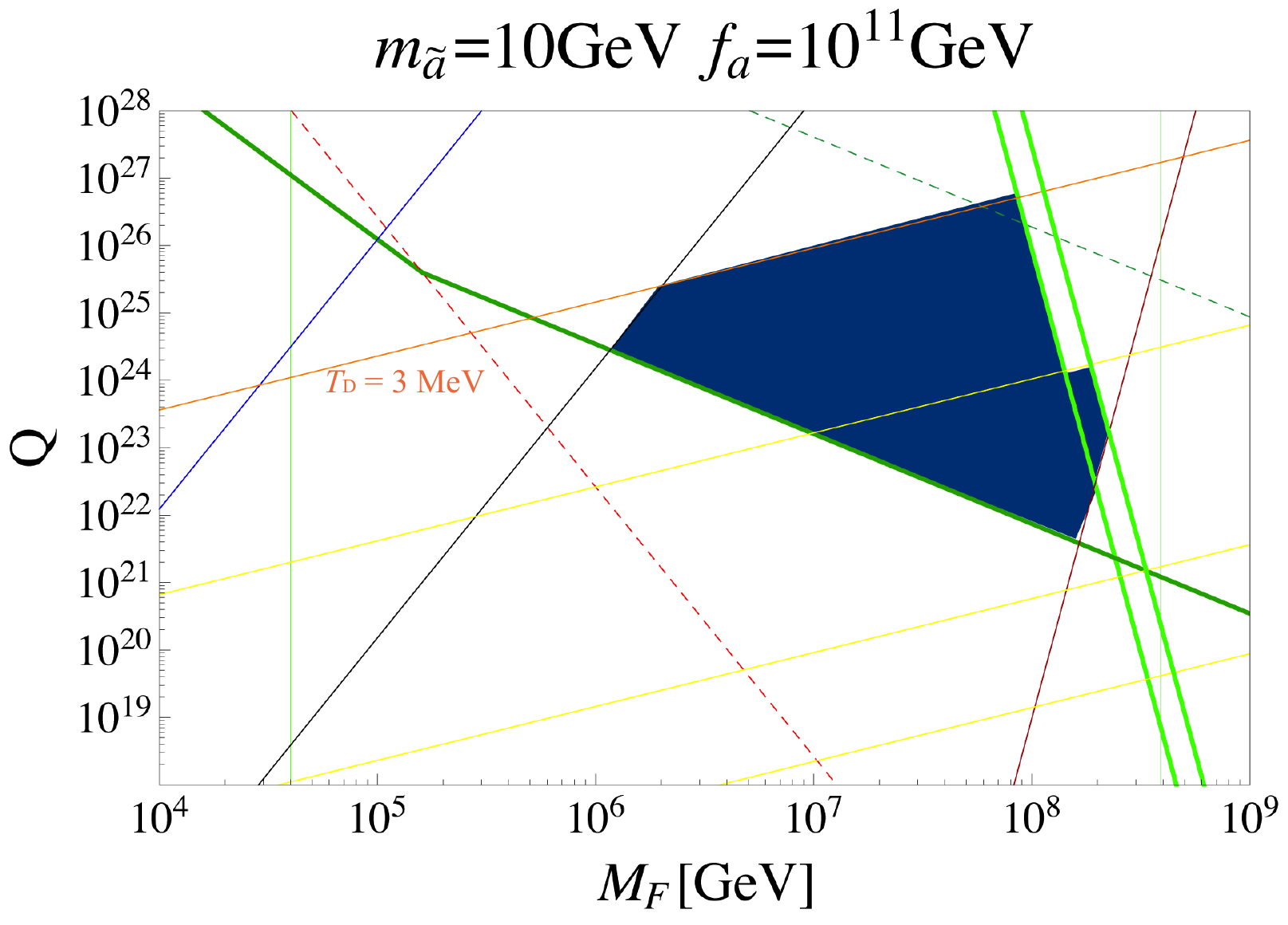} &
\includegraphics[width=85mm]{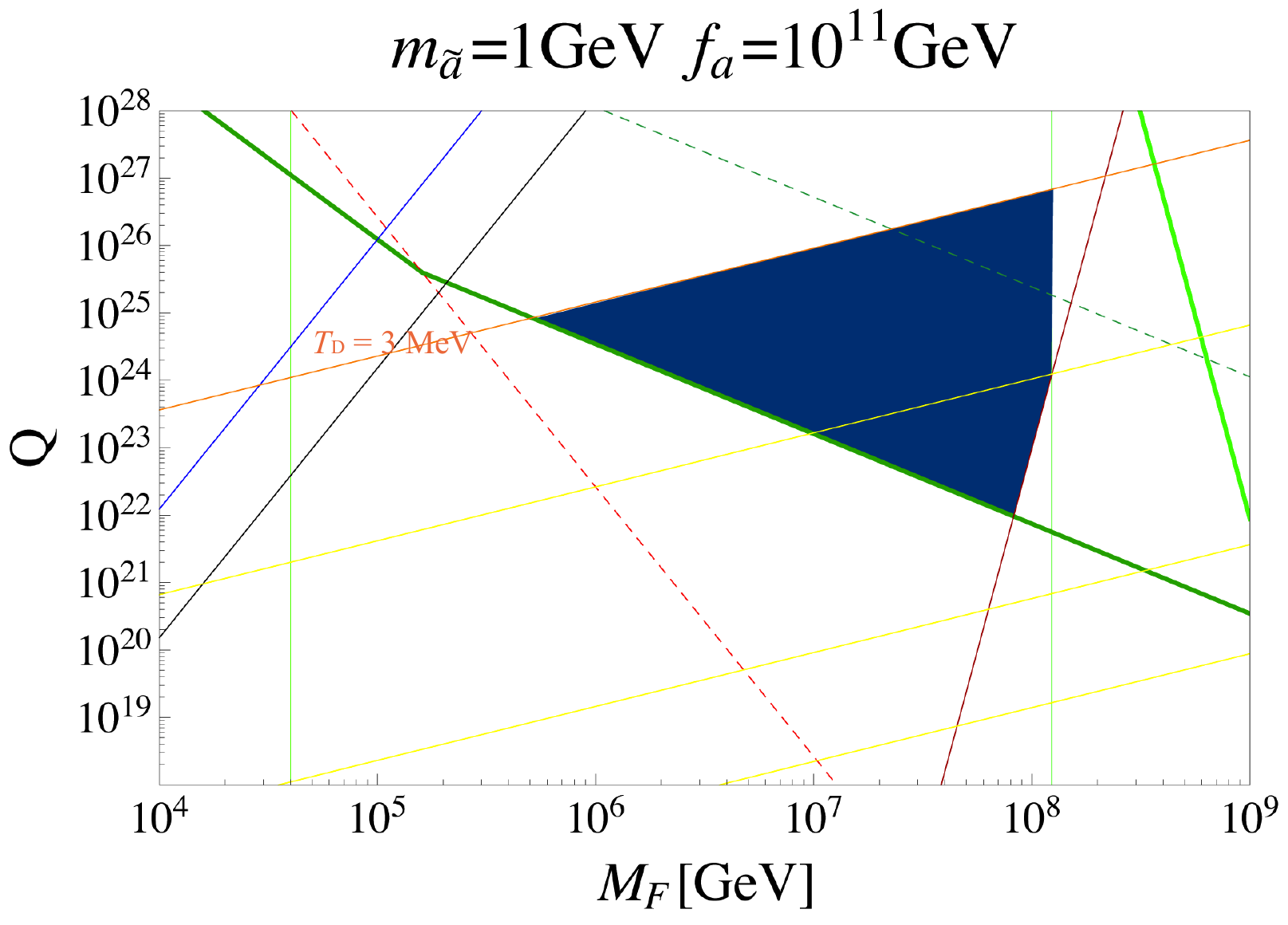}\\
\includegraphics[width=85mm]{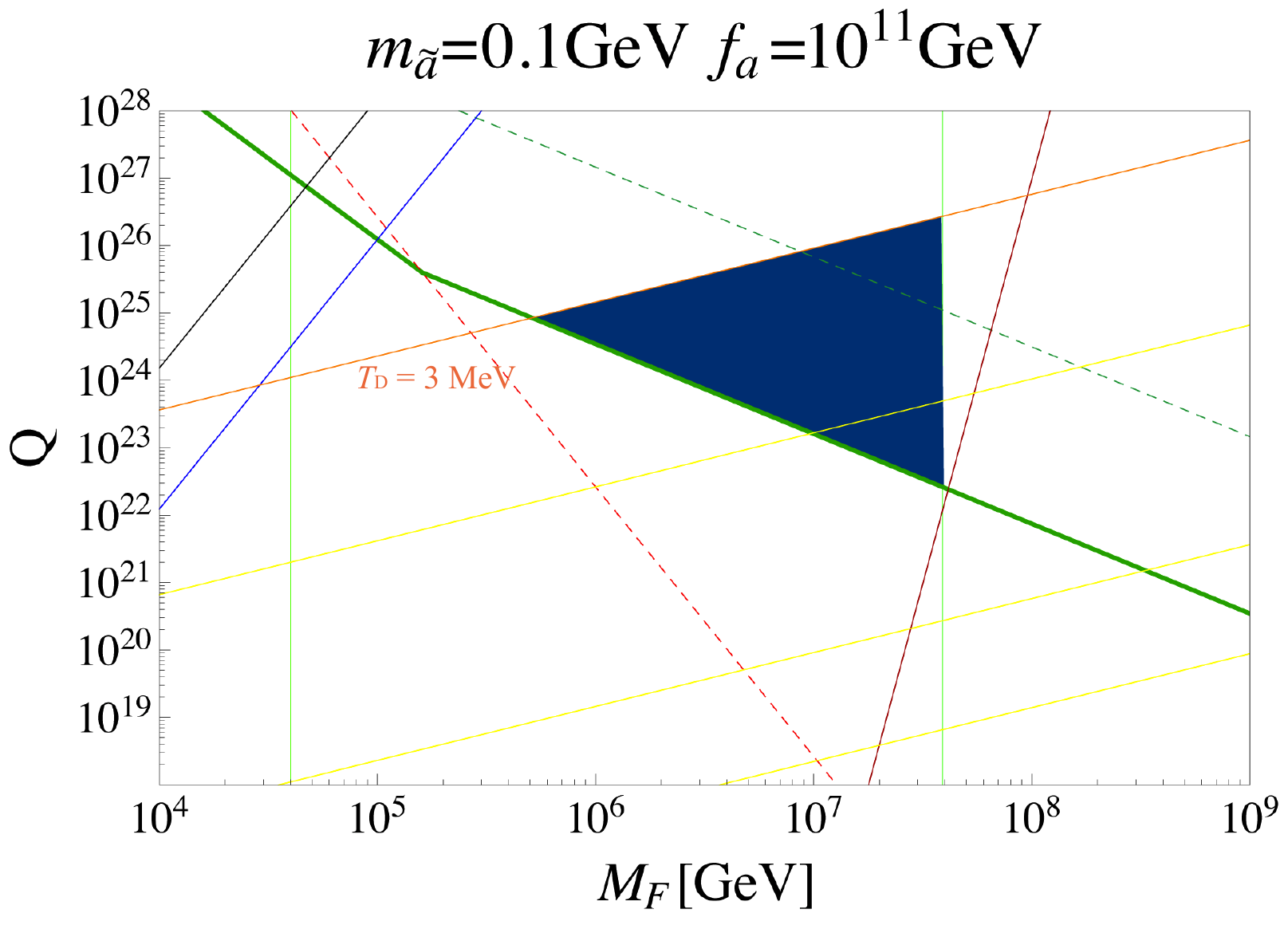} &
\includegraphics[width=85mm]{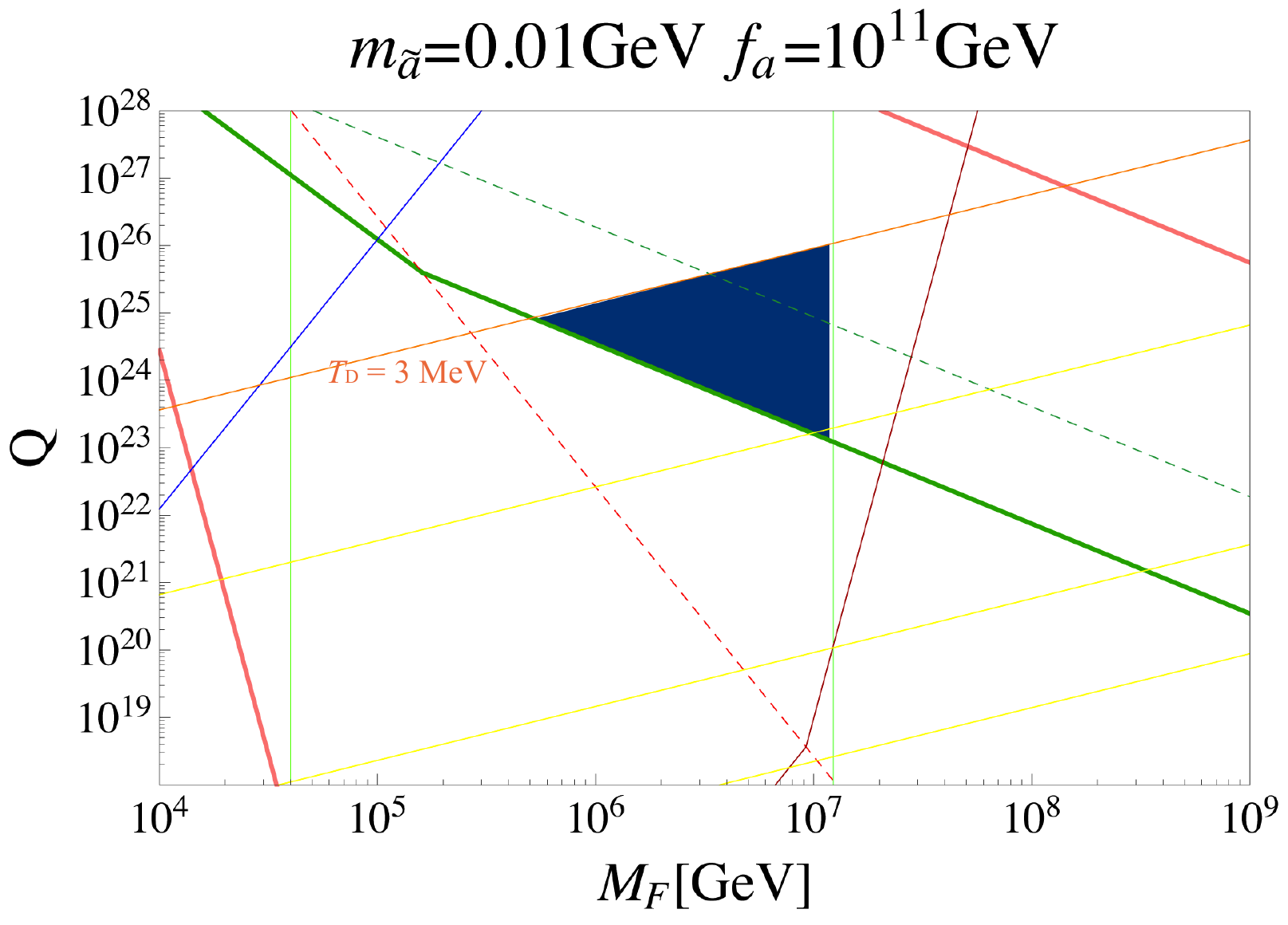}\\
\multicolumn{2}{c}{
\includegraphics[width=85mm]{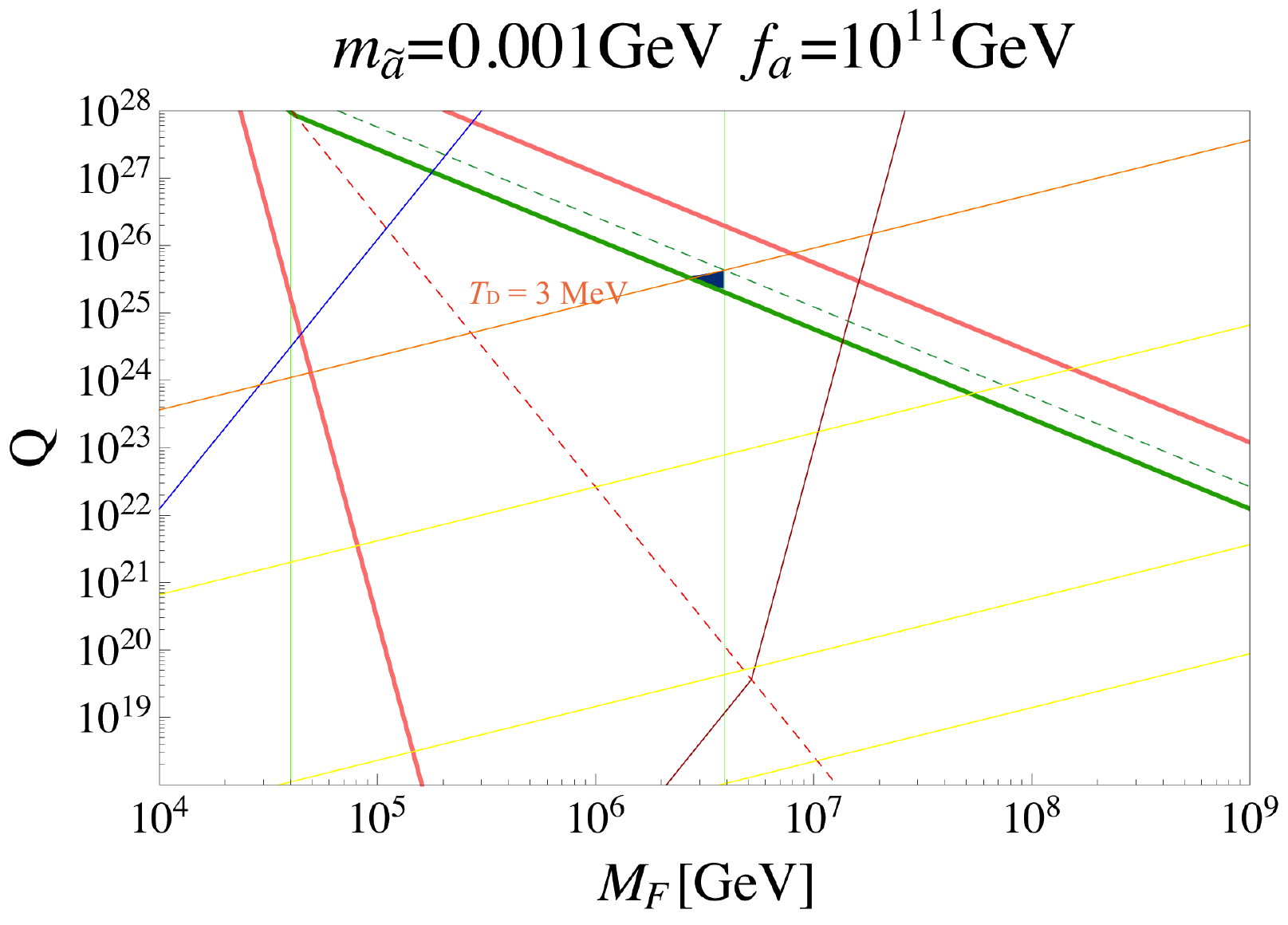}}
\end{tabular}
\caption{Allowed regions for $f_a=10^{11}$~GeV in both the bino and stau MLSP cases for KSVZ models.}
\label{Qpara11}
\end{center}
\end{figure}

\begin{figure}
\begin{center}
\begin{tabular}{cc}
\includegraphics[width=85mm]{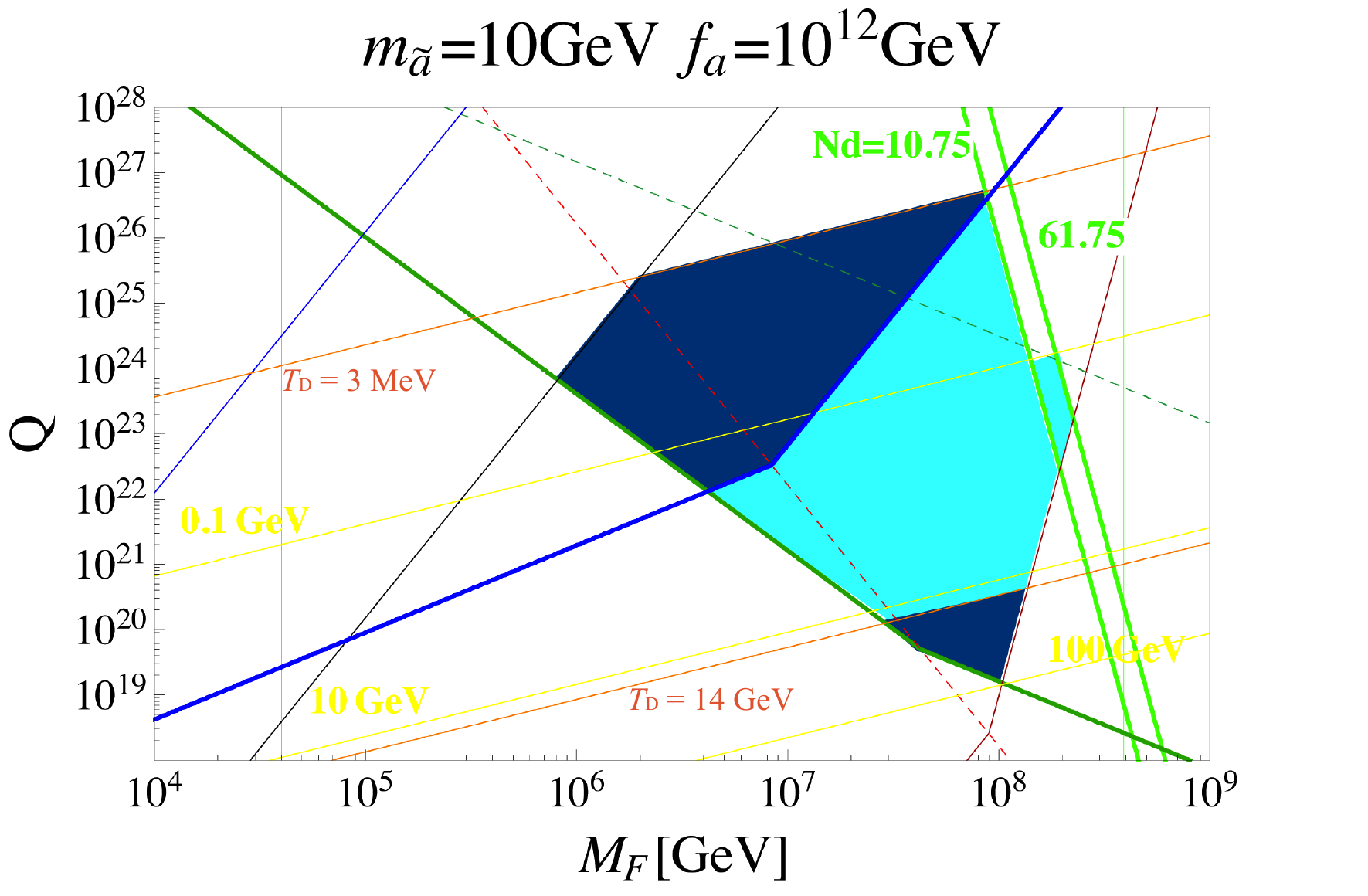} &
\includegraphics[width=85mm]{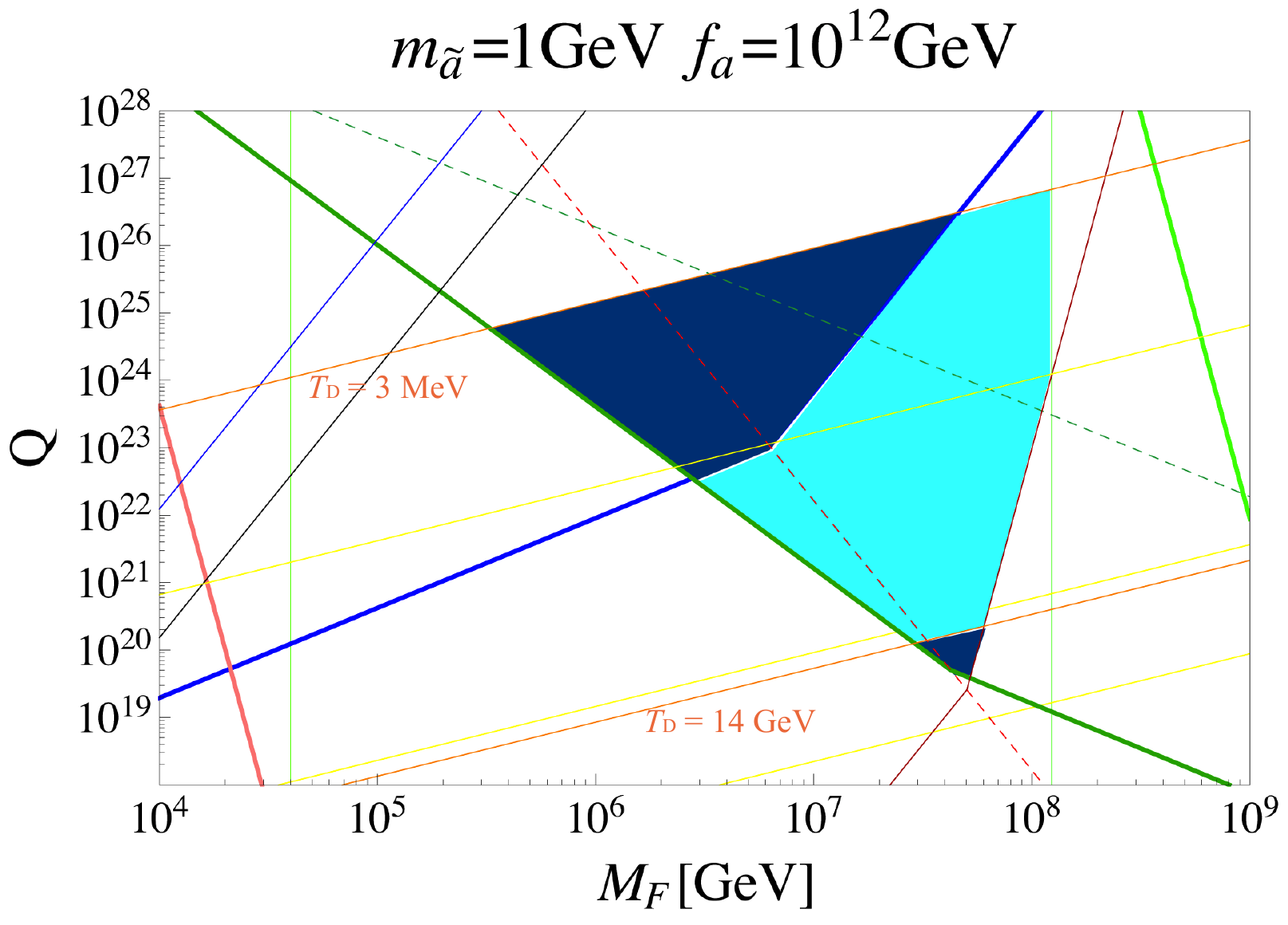}\\
\includegraphics[width=85mm]{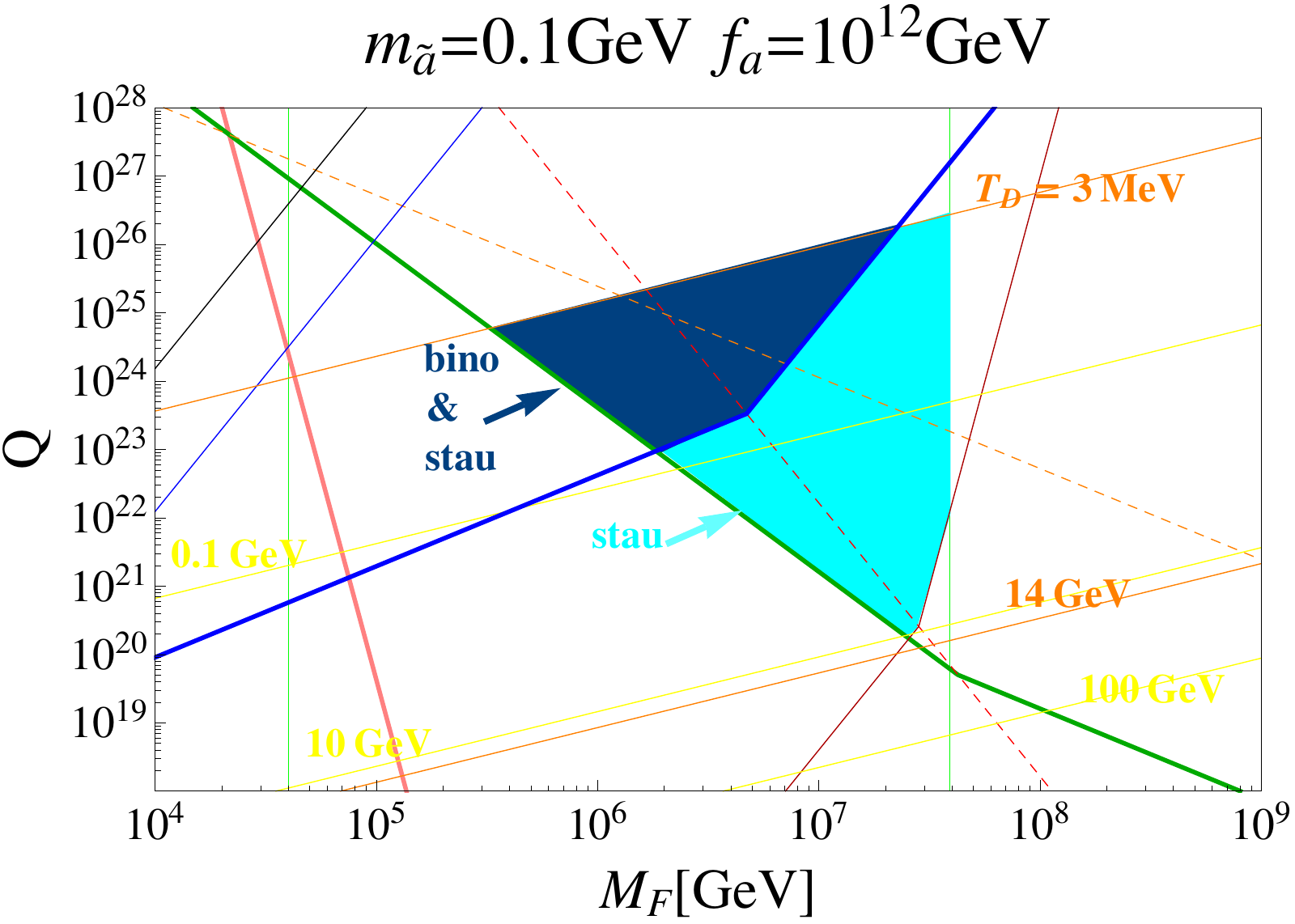} &
\includegraphics[width=85mm]{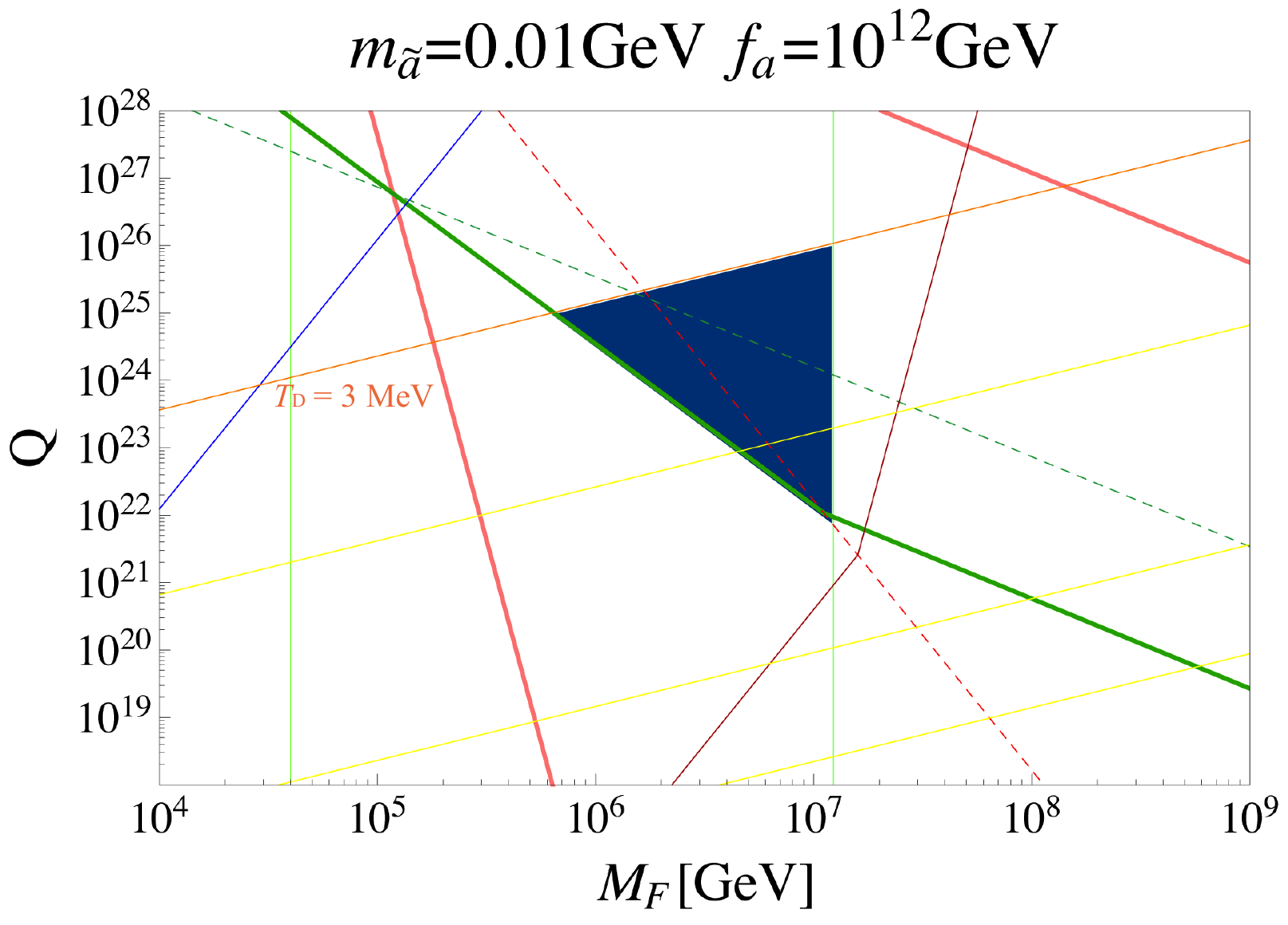}\\
\multicolumn{2}{c}{
\includegraphics[width=85mm]{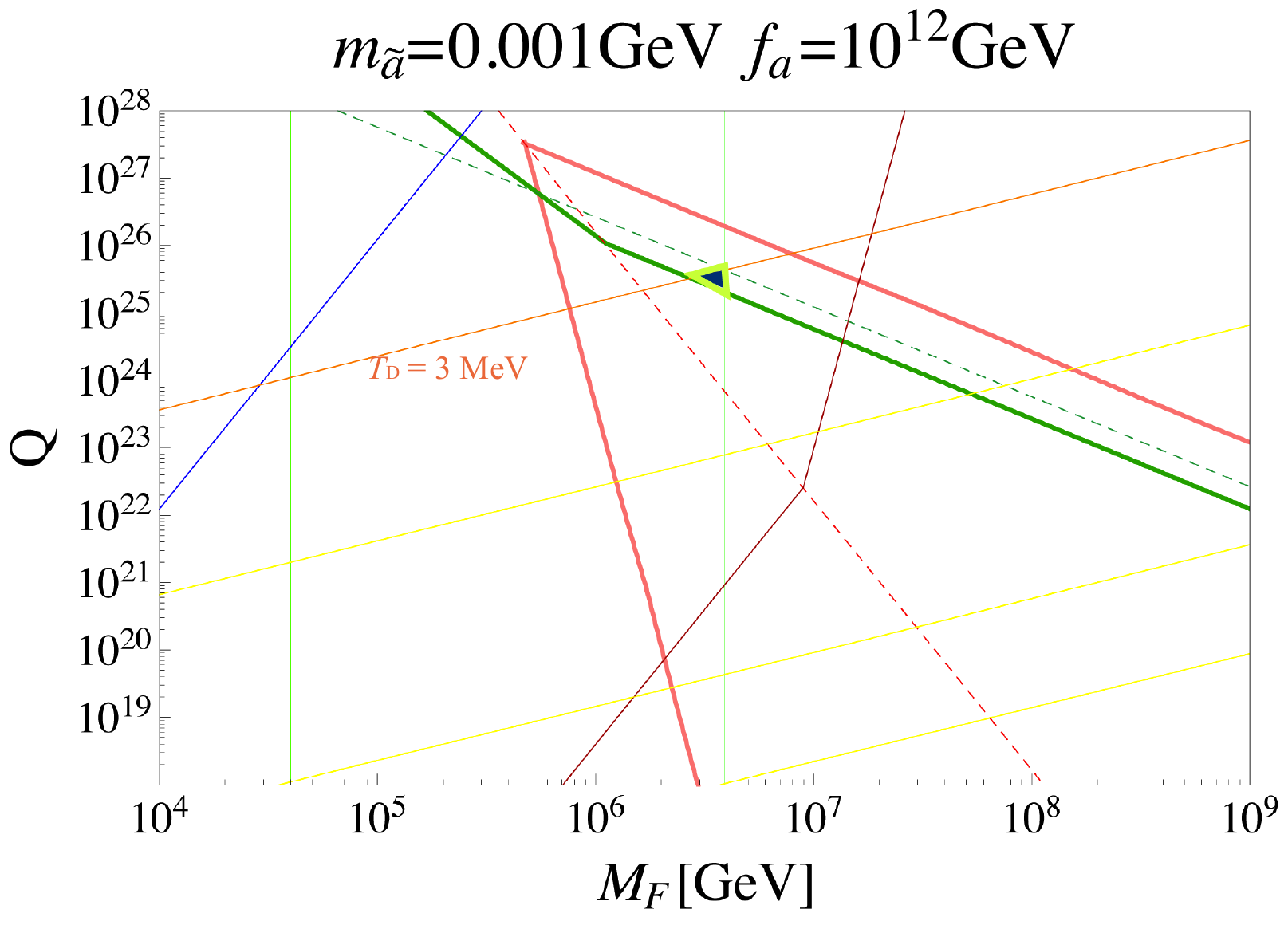}}
\end{tabular}
\caption{Allowed regions for $f_a=10^{12}$~GeV for KSVZ models. Dark blue areas are allowed for
both the bino and stau MLSP cases, while it works only the stau MLSP in cyan areas. DFSZ case is
allowed only for $m_{\tilde{a}}=0.001$~GeV.}
\label{Qpara12}
\end{center}
\end{figure}

\begin{figure}
\begin{center}
\begin{tabular}{cc}
\includegraphics[width=85mm]{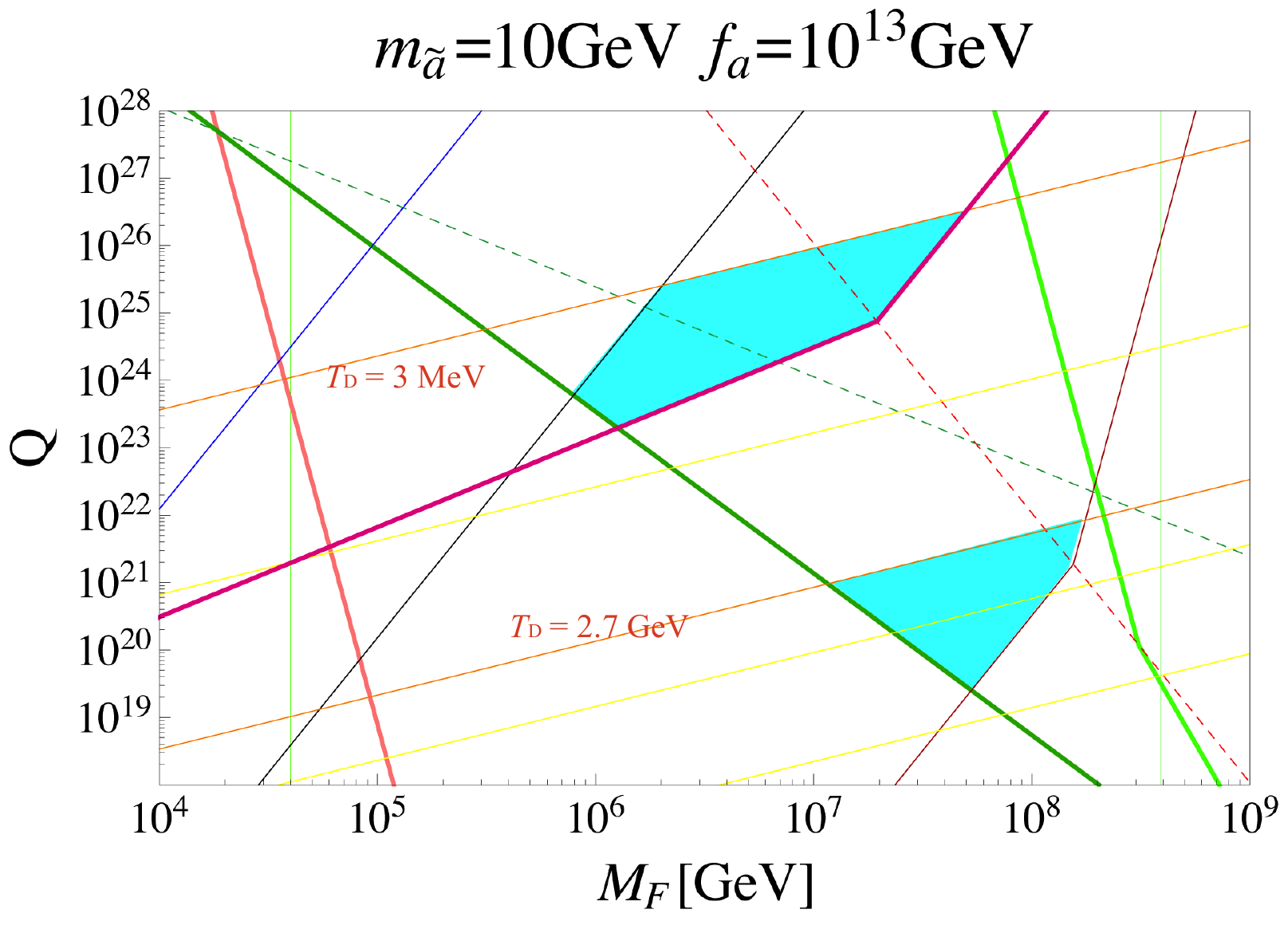} &
\includegraphics[width=85mm]{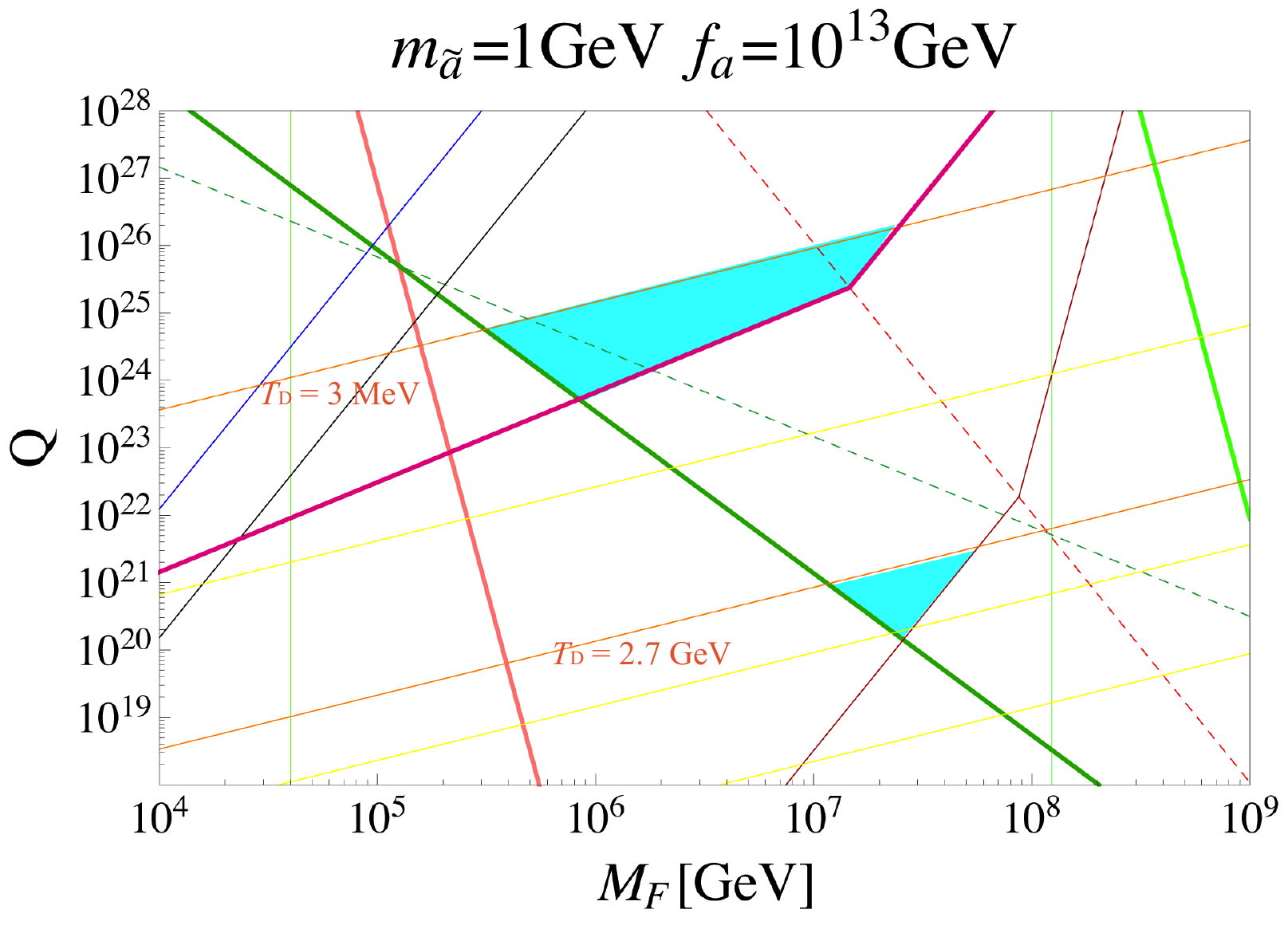}\\
\includegraphics[width=85mm]{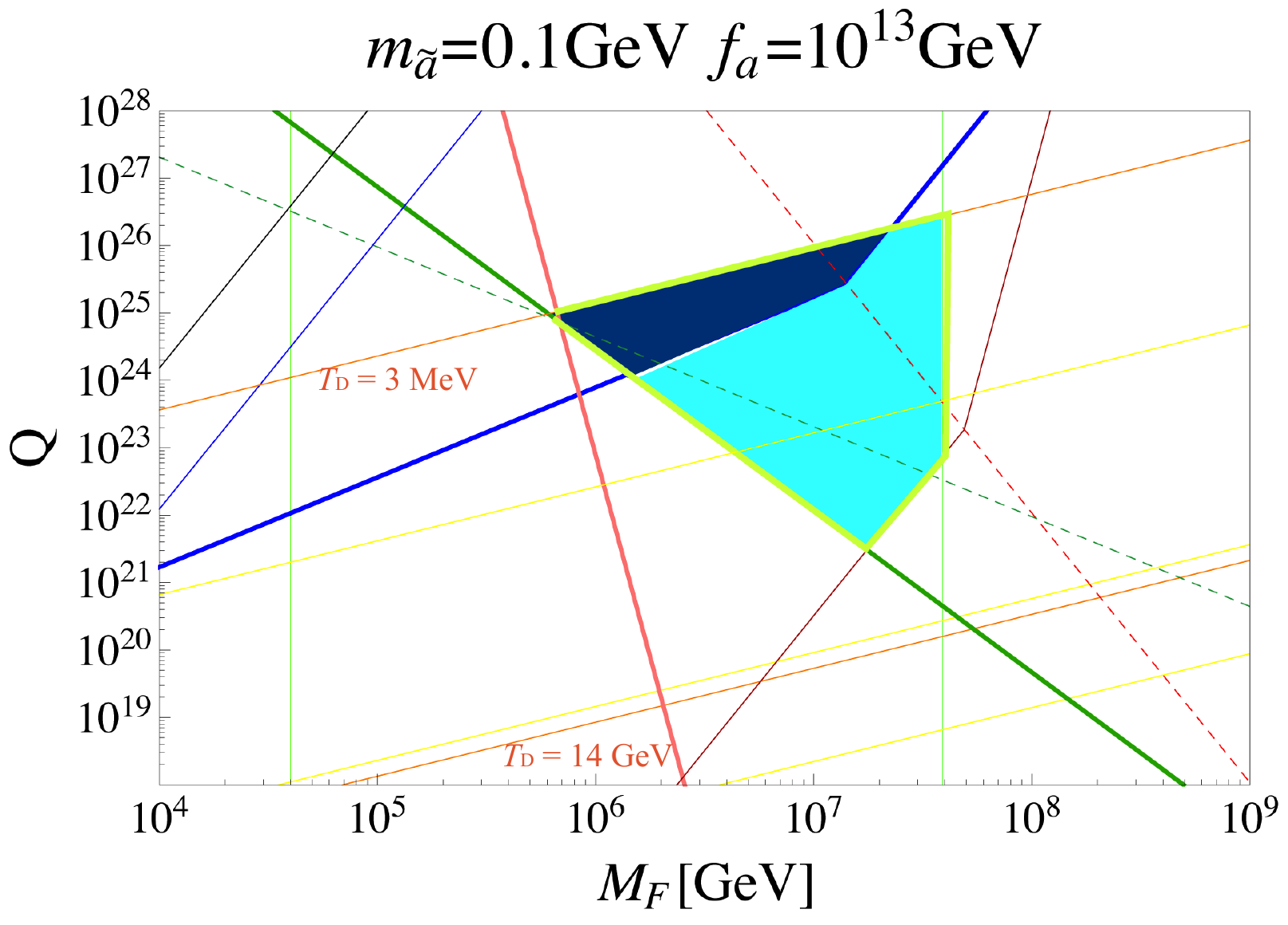} &
\includegraphics[width=85mm]{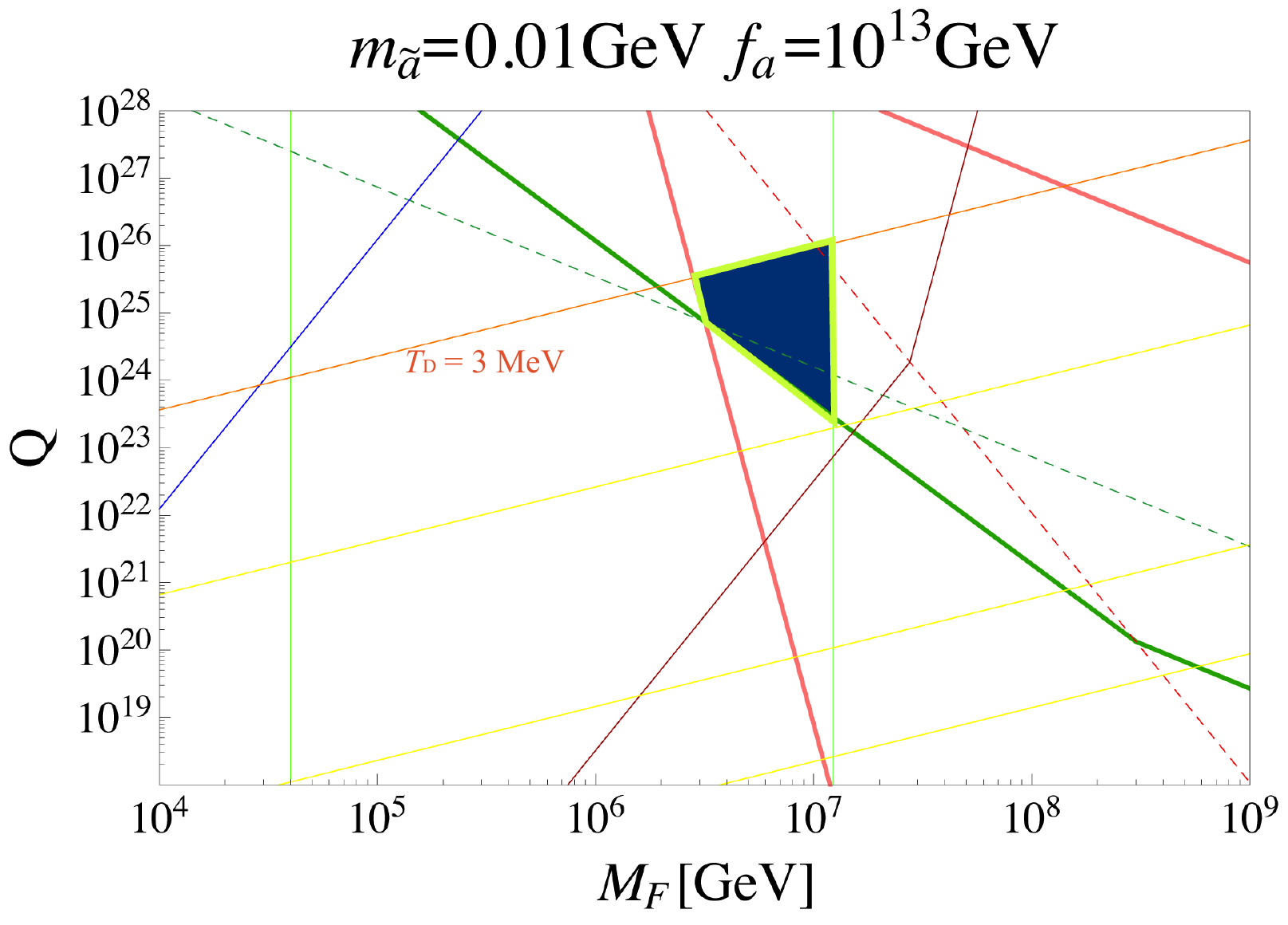}\\
\end{tabular}
\caption{Allowed regions for $f_a=10^{13}$~GeV for KSVZ models. Dark blue areas are allowed for
both the bino and stau MLSP cases, while it works only the stau MLSP in cyan areas. DFSZ case is
allowed only for $m_{\tilde{a}}=0.1$ and 0.1~GeV.}
\label{Qpara13}
\end{center}
\end{figure}

\begin{figure}
\begin{center}
\includegraphics[width=85mm]{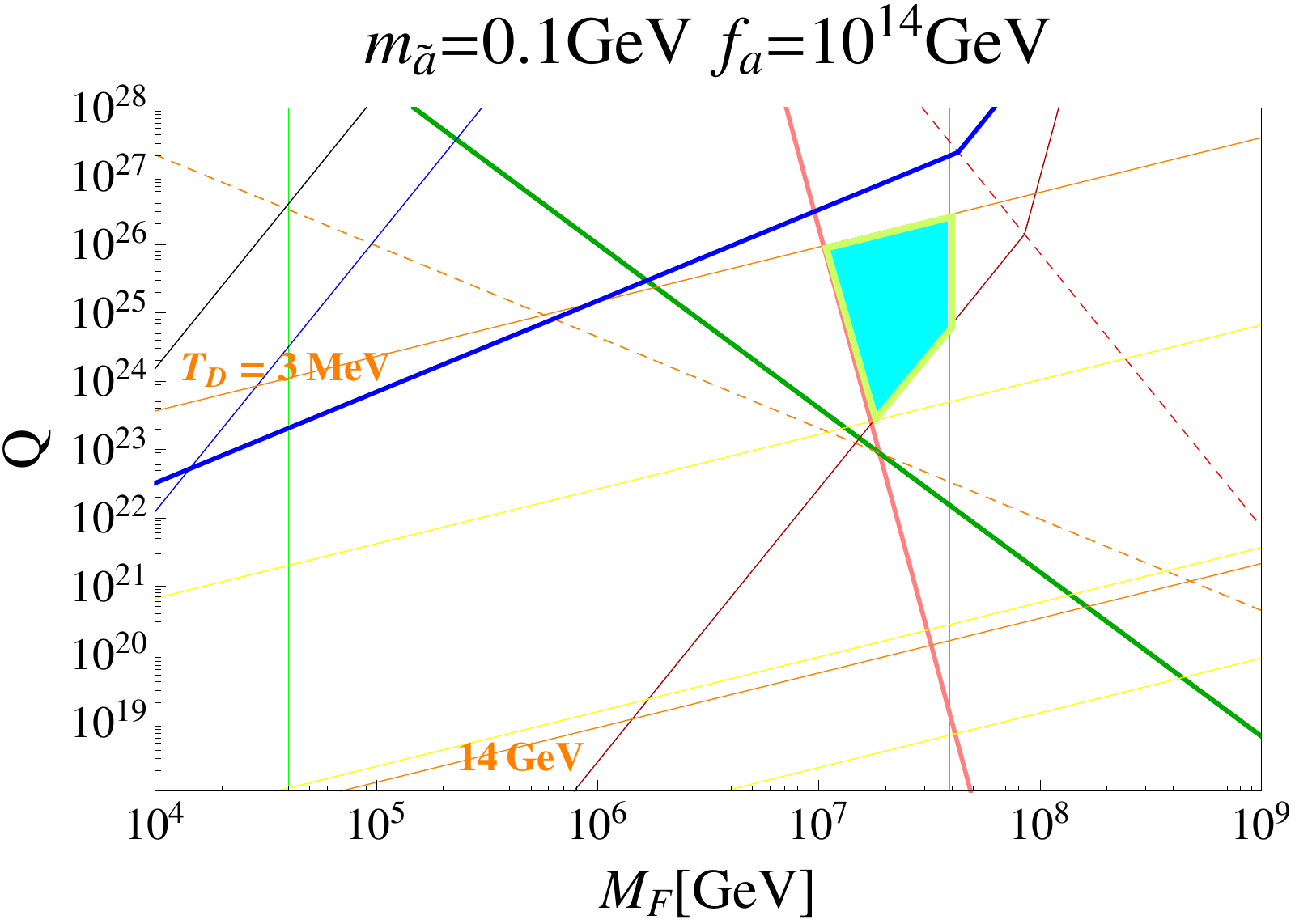}
\caption{Allowed regions for $f_a=10^{14}$~GeV for both KSVZ and DFSZ models. It works only for the 
stau MLSP.}
\label{Qpara14}
\end{center}
\end{figure}

We show Eqs.(\ref{Qsat2QD}) and (\ref{Qunsat2QD}) as thick pink lines, Eqs.(\ref{Qsat2NQDU}) and 
(\ref{Qunsat2NQDU}) as thick green lines, and Eqs.(\ref{Qsat2NQDL}) - (\ref{Qunsat2NQDL}) 
as thick dark green lines for $f_a=10^{11} - 10^{14}$~GeV respectively in 
Figs.~\ref{Qpara11} $-$~\ref{Qpara14}. Here we display only the figures 
with those $f_a$ and $m_{\tilde{a}}$ that our scenario works. Saturation and unsaturation are divided 
by the condition $B_{\tilde{a}}^{(\rm{sat})} = B_{\tilde{a}}^{(\rm{unsat})}$ (See Eqs.(\ref{branchsat}) and 
(\ref{branchunsat})). It is rewritten as
\begin{equation}
Q = 3.0 \times 10^{31} \left( \frac{f_a}{10^{12}\, {\rm GeV}}\right)^4
\left( \log\left(\frac{f_{a}}{10^3 \, {\rm GeV}}\right) \right)^{-4}
\left( \frac{M_F}{10^6 \, {\rm GeV}}\right)^{-4}
\left( \frac{N_q}{18}\right)^{-2} \left( \frac{\zeta}{2.5} \right)^{-4},
\end{equation}
shown by red dashed lines in the figures. Whether $Q$-ball dominates or not at the decay 
is determined by the condition if Eq.(\ref{QD/NQD}) is larger or smaller than unity. We thus have
the line
\begin{eqnarray}
Q & = & 4.4 \times 10^{23} 
\left( \frac{\beta}{6\times 10^{-5}}\right)^{2/3} 
\left( \frac{T_{\rm RH}}{10^7 \, {\rm GeV}}\right)^{-8/9}
\left( \frac{M_F}{10^6 \, {\rm GeV}}\right)^{-4/3} 
\nonumber \\ & & \hspace{50mm} \times
\left( \frac{N_q}{18}\right)^{4/9} 
\left( \frac{N_d}{10.75}\right)^{-2/9} \left( \frac{\zeta}{2.5} \right)^{-4/9},
\end{eqnarray}
to separate the parameter space, where we show this for $T_{\rm RH} =T_{\rm RH,max}$ 
in dark green dashed lines in the figures.


There are four conditions for the $Q$-ball decay to be satisfied. The $Q$ ball is kinematically allowed 
to decay into (a) axinos and (b) nucleons. (c) Branching of the decay into gravitinos should be 
suppressed compared to the decay into axinos. (d) $Q$-ball decay must complete before the BBN, which
we assume to be $T_{\rm D} > 3$~MeV. The condition (a) and (b) can be written as
\begin{equation}
Q \lesssim 1.5 \times 10^{28} 
\left( \frac{M_F}{10^6 \,\mathrm{GeV}}\right)^4 
\left( \frac{\omega_{Q}}{\mathrm{GeV}}\right)^{-4}
\left(\frac{\zeta}{2.5}\right)^{4},
\end{equation}
with $\omega_{Q}=m_{\tilde{a}}$ (black lines) and $\omega_{Q}=bm_N$ (blue lines), respectively.
The condition (c) is rephrased from Eq.(\ref{branchingaxinotogravitino}) as
\begin{eqnarray}
& & 
Q_{\rm sat} \gtrsim 1.0 \times 10^7
\left( \frac{m_{\tilde{a}}}{10\,\rm MeV}\right)^{-4}
\left( \frac{N_q}{18}\right)^2
\left( \frac{M_F}{10^6\, \mathrm{GeV}}\right)^{12}
\left( \frac{\zeta}{2.5}\right)^{12}
\left( \frac{m_{\tilde{g}}}{1\,\rm TeV}\right)^{-4}, \\
& &
Q_{\rm unsat} \gtrsim 1.7 \times 10^{19}
\left( \frac{m_{\tilde{a}}}{10\,\rm MeV}\right)^{-2}
\left( \frac{f_a}{10^{12}\, \mathrm{GeV}}\right)^2
\left( \log\left(\frac{f_{a}}{10^3 \, {\rm GeV}}\right) \right)^{-2}
\left( \frac{M_F}{10^6\, \mathrm{GeV}}\right)^4
\nonumber \\ & & \hspace{100mm} \times 
\left( \frac{\zeta}{2.5}\right)^4
\left( \frac{m_{\tilde{g}}}{1\,\rm TeV}\right)^{-2}, 
\end{eqnarray}
for the saturated and unsaturated cases, respectively. They are denoted by dark red lines in the figures.
The condition (d) is given by, from Eq.(\ref{td}), 
\begin{equation}
Q \lesssim 1.5 \times 10^{25} 
\left( \frac{M_F}{10^6\, \mathrm{GeV}}\right)^{4/5}
\left( \frac{T_{\rm D}}{\mathrm{3\, MeV}}\right)^{-8/5} 
\left( \frac{N_q}{18}\right)^{4/5}
\left( \frac{N_d}{10.75}\right)^{-2/5}
\left( \frac{\zeta}{2.5}\right)^{4/5},
\end{equation}
displayed by orange lines with $T_{\rm D} =3$~MeV in the figures. 
We also plot $T_{\rm D}=0.1$, 10, and 100 GeV in yellow lines.

In addition, we must have $\epsilon < 1$ and $M_F$-limit (Eq.(\ref{Mf})). We can easily see that
the former condition is always satisfied with parameters in our successful scenario, while the latter
only restricts $M_F$ as
\begin{equation}
4\times 10^4 \, {\rm GeV} \lesssim M_F 
\lesssim 1.2 \times 10^7 \, {\rm GeV} \left( \frac{m_{\tilde{a}}}{10\,\rm MeV}\right)^{1/2},
\end{equation}
shown as light green lines.

Finally, the parameter space is constrained by the upper limit on the MLSP abundance 
(\ref{BBNconbino}) or (\ref{BBNconstau}), since $Q$ can be expressed in terms of the MLSP abundance. 
Using Eq.(\ref{rhomfqsat}) for the saturated case, and Eq.(\ref{rhomfqunsat}) for the unsaturated case, 
we respectively obtain
\begin{eqnarray}
Q_{\rm sat} & \simeq &
2.8 \times 10^{13} 
\left(\frac{Y_b}{10^{-10}}\right)
\left( \frac{\rho_{\rm MLSP}/s}{\rm GeV}\right)^{-1} 
\left( \frac{m_{\tilde{a}}}{10\,\rm MeV}\right)^{-1} 
\left( \frac{m_{\rm MLSP}}{300\, {\rm GeV}}\right)^{-3} 
\left( \frac{M_F}{10^6\, {\rm GeV}}\right)^{4}
\left( \frac{\zeta}{2.5}\right)^{4},
\nonumber \\ 
\label{Qsat1} & & \\
\label{Qunsat1}
Q_{\rm unsat} & \simeq &
2.9 \times 10^{19} 
\left(\frac{Y_b}{10^{-10}}\right)^{2/3}
\left( \frac{\rho_{\rm MLSP}/s}{\rm GeV}\right)^{-2/3} 
\left( \frac{m_{\tilde{a}}}{10\,\rm MeV}\right)^{-2/3}
\left( \frac{m_{\rm MLSP}}{300 \,{\rm GeV}}\right)^{-2}
\nonumber \\ & & \hspace{5 mm} \times 
\left( \frac{f_a}{10^{12}\,{\rm GeV}}\right)^{4/3}
\left( \log\left(\frac{f_a}{10^3 \,{\rm GeV}}\right)\right)^{-4/3}
\left( \frac{M_F}{10^6\, {\rm GeV}}\right)^{4/3}
\left( \frac{N_q}{18}\right)^{-2/3}
\left( \frac{\zeta}{2.5}\right)^{4/3}.
\end{eqnarray}
Thus, we obtain the lower bound on $Q$ inserting Eq.(\ref{BBNconbino}) for the bino MLSP or 
Eq.(\ref{BBNconstau}) for the stau MLSP, displayed in thick blue and magenta lines, respectively.

We hatch the allowed regions in the parameter space ($Q$, $M_F$) in these figures: the cyan areas
are for the stau MLSP and the dark blue areas are for both the bino and stau MLSP for KSVZ case.
The allowed regions for DFSZ case are surrounded by thick yellow lines, appeared only for 
($f_a$, $m_{\tilde{a}}$)$=(10^{12}$~GeV, 0.001~GeV), ($10^{13}$~GeV, 0.01~GeV), 
($10^{13}$~GeV, 0.1~GeV), and ($10^{14}$~GeV, 0.1~GeV).
In this region, the axino overproduction by the higgsino decay can be avoided.
Note that typically for the regions $T_{\rm D} > 14$~GeV for the bino MLSP case and 
$T_{\rm D}>2.7$~GeV for the stau case, there is no BBN constraints due to 
the large annihilations [Eqs.~(\ref{annihilationbino}), (\ref{annihilationstau})]. 
There are thus isolated allowed regions in the lower part of the parameter space.
We see that our scenario works in rather wide parameter ranges, 
typically for $Q=10^{20} - 10^{26}$ and $M_F=10^6 - 10^8$ GeV, 
and for $f_a= 10^{11} - 10^{14}$~GeV with $m_{\tilde{a}}=0.01 - 10$~GeV.

\section{Summary} 
We have investigated the $Q$-ball scenario in the gauge-mediated SUSY breaking 
model where the $Q$ ball decays into axinos and nucleons, providing simultaneously 
dark matter and the baryon asymmetry of the universe. The branching of the $Q$-ball 
decay into axinos is typically small, but the decay tends to be saturated for smaller $f_a$.
This is in contrast to the branching into the gravitino which is always unsaturated. 
The branching into the gravitino is mostly much smaller than that of the decay into axinos.
SUSY particles in the MSSM, e.g., the MLSPs, could be produced by the $Q$-ball decay,
but the MLSPs annihilate afterwards and their abundance becomes small enough to
evade the BBN constraints.

The successful scenario resides in the regions typically for
$m_{\tilde{a}}=0.01 - 10$~GeV and $f_a = 10^{11} - 10^{14}$~GeV in the KSVZ model, 
while the small $f_a$ region ($f_a \lesssim 10^{12}$~GeV) and the large $m_{\tilde{a}}$ region 
($m_{\tilde{a}} \lesssim 1$~GeV) are excluded in the DFSZ model. This can be realized in 
the $Q$-ball parameters as $Q=10^{20}-10^{26}$ and $M_F=10^6 - 10^8$~GeV.

Finally, we comment on the free streaming of the axino. Because of the rather large
kinetic energy of the axinos emitted from $Q$ balls, the free streaming might affect
the structure formation of the universe. For the parameters of the successful
scenario, we have a shorter free streaming length than $\sim$ Mpc. Thus, we can safely
neglect such effects.

\appendix
\makeatletter
  \renewcommand{\theequation}{%
        \thesection.\arabic{equation}}
 \@addtoreset{equation}{section}
\makeatother

\section{Decay rates of the neutralinos into axinos and gravitinos}
Now we evaluate the MLSP decay into the axinos and the gravitinos. 
Firstly for the bino MLSP case, the decay rate of the bino into the axino is given by \cite{CKKR}
\begin{eqnarray}
\label{GMLSP}
\Gamma_{\mathrm{MLSP}\rightarrow\tilde{a}}^{\tilde{B}} 
& = & \frac{\alpha_{\rm em}^2C^2}{128\pi^3\cos^4{\theta_{W}}}
\frac{m_{\tilde{B}}^3}{f_a^2}\left( 1-\frac{m_{\tilde{a}}^2}{m_{\tilde{B}}^2}\right)^3, 
\end{eqnarray}
where $\alpha_{\rm em}=1/137$ is the electromagnetic coupling strength and $\theta_{\rm W}$ is the 
Weinberg angle, $C$ is a model dependent parameter which we take $C=1$. On the other hand, 
the decay rate of the bino into the gravitino is given by \cite{Feng:2004zuFeng:2004mt}
\begin{eqnarray}
\label{binogravitino}
\Gamma_{\mathrm{MLSP}\rightarrow3/2}^{\tilde{B}} 
&=& \frac{\cos^2{\theta_{W}}}{48\pi M_{\rm P}^2} \frac{m^5_{\tilde{B}}}{m^2_{3/2}}
\left( 1-\frac{m^2_{3/2}}{m^2_{\tilde{B}}}\right)^3\left( 1+3\frac{m^2_{3/2}}{m^2_{\tilde{B}}}\right ).
\end{eqnarray}
Next we consider the stau MLSP case. The decay rate of stau into axino is given by \cite{Freitas:2011fx}
\begin{equation}
\label{stauaxino}
\Gamma_{\mathrm{MLSP}\rightarrow \tilde{a}}^{\tilde{\tau}} 
= \frac{m_{\tilde{\tau}}}{16\pi}\left( 1-\frac{m^2_{\tilde{a}}}{m^2_{\tilde{\tau}}}\right)^2 |A|^2, \\
\end{equation}
where
\begin{equation}
A = \frac{3\alpha_{\rm em}^2e_{Q}^2}{8\pi^2\cos^4{\theta_{\rm W}}}
\frac{\sqrt{2}m_{\tilde{\tau}}}{f_a}3\frac{m_{\tilde{B}}}{m_{\tilde{\tau}}}
\log{\left(\frac{y^2 f_{a}^2}{2m_{\tilde{\tau}}^2}\right)}.
\end{equation}
Here we assume the bino mass as $1\,\mathrm{TeV}$ (only for the stau MLSP case), 
$e_{Q}^2 = C/6$ and $y$ is a model dependent parameter which we set 1 here. 
On the other hand, the decay rate of the stau into the gravitino is expressed as \cite{Kawasaki:2007xb}
\begin{eqnarray}
\label{staugravitino}
\Gamma_{\mathrm{MLSP}\rightarrow 3/2}^{\tilde{\tau}} 
&=& \frac{m_{\tilde{\tau}}^5}{48\pi m_{3/2}^2M_{\rm P}^2}\left(1-\frac{m_{3/2}^2}{m^2_{\tilde{\tau}}}\right)^4.
\end{eqnarray}

\section{The thermally produced axinos in two axion models}
We want to estimate the abundance of the axinos thermally produced by scattering
processes. In this paper, we adopt the results of Ref.\cite{HTLpaper}. We notice that it
is only valid for small coupling regime, and there may be ambiguities of about an order of
the magnitude \cite{HTLpaper,Strumia,Choi:2011yf}, or even some controversies on the
estimate of the axino-gluon-gluino coupling \cite{BCI}. 

The axino production from the scattering via the axino-gluino-gluon interaction can 
be expressed in a gauge invariant way. The axino yield, 
$Y_{\tilde{a}}=\frac{n_{\tilde{a}}}{s}$, at present can be obtained by
\begin{equation}
Y_{\tilde{a}}  \simeq 
\frac{C_{\tilde{a}}(T_{\mathrm{RH}})}{s(T_{\mathrm{RH}})H(T_{\mathrm{RH}})},
\end{equation}
with the collision term for SU($N$) \cite{HTLpaper}
\begin{eqnarray}
C_{\tilde{a}}(T) \simeq \frac{(N^2-1)}{f_a^2}\frac{3\zeta(3)g^6T^6}{4096\pi^7}
\left[\log{\left(\frac{1.647T^2}{m_g^2}\right)}\left(N+n_f \right)+0.4336n_f\right],
\end{eqnarray}
where $g$ is a coupling constant of SU($N$) and $n_f$ is a number of SU($N$) 
multiplet and anti-multiplet, and $m_g=gT\sqrt{\frac{N+n_f}{6}}$ is the thermal 
SU($N$)-gaugino mass. Here, we use the Hubble parameter 
$H(T)=\sqrt{\frac{g_*(T)\pi^2}{90}}\frac{T^2}{M_{\rm P}}$ 
and the entropy density $s(T)=\frac{2\pi^2}{45}g_*(T)T^3$, where $g_*$ is the number of 
effectively massless degrees of freedom and we use $g_* = 228.75$. 
Then the axino density parameter is estimated as
\begin{eqnarray}
\label{thaa}
\Omega_{\tilde{a}}^{\mathrm{TH}}h^2 
&=& m_{\tilde{a}}Y_{\tilde{a}}\frac{s(T_{0})h^2}{\rho_c}, \nonumber\\
&\simeq& 7.7\times10^{-4}g^6(N^2-1)
\left(\frac{m_{\tilde{a}}}{\rm GeV}\right)
\left( \frac{f_{a}}{10^{14}\,\mathrm{GeV}}\right)^{-2}
\left(\frac{T_{\mathrm{RH}}}{10^7\,\mathrm{GeV}}\right) \nonumber\\
& & \hspace{15mm} \times \left[
\log{\left( \frac{3.144}{g\sqrt{N+n_f}}\right)}
\left(N+n_f \right)+0.2168n_f
\right],
\end{eqnarray}
where $\rho_c$ is the present critical density. 

When the SU(3) anomaly term is present as in the KSVZ model, Eq.(\ref{thaa}) 
can be rewritten as
\begin{equation}
\Omega_{\tilde{a}}^{\mathrm{TH}(\mathrm{KSVZ})}h^2 
\simeq  1.0 
\left( \frac{m_{\tilde{a}}}{10\,\rm MeV}\right) 
\left( \frac{f_{a}}{10^{12}\,\mathrm{GeV}}\right)^{-2} 
\left( \frac{T_{\rm RH}}{10^7\,\mathrm{GeV}}\right),
\label{appOmegaKSVZ}
\end{equation}
where $g$ is the coupling constant of SU(3)$_C$, and we use $g = 0.983$, the value 
at $10^6\,\mathrm{GeV}$, and $n_f=6$ in the second equality. When the SU(3) anomaly
term is absent as in the case for the high temperature regime in the DFSZ model, we need
to consider the SU(2)$_L$ anomaly term \cite{Choi:2011yf}. Eq.(\ref{thaa}) is then given as 
\begin{equation}
\label{appOmegaDFSZ2}
\Omega_{\tilde{a}}^{\mathrm{TH}(\mathrm{DFSZ})}h^2 
= 0.1\left( \frac{m_{\tilde{a}}}{10\rm MeV}\right)
\left(\frac{f_{a}}{10^{12}\,\mathrm{GeV}}\right)^{-2}
\left(\frac{T_{\mathrm{RH}}}{10^7\,\mathrm{GeV}}\right),
\end{equation}
where $N=2$, $n_f=14$ and $g=0.663$, estimated at $10^6\,$GeV, are used. 

In the DFSZ model, there also exists a tree-level axino-Higgs-higgsino coupling which 
contributes to the thermally produced axinos by the higgsino decay whose decay 
rate is given by \cite{Chun,Choi:2011yf}
\begin{equation}
\Gamma_{\tilde{h}} \simeq c_H^2\left(\frac{\mu}{f_{a}}\right)^2\frac{m_{\tilde{h}}}{16\pi},
\end{equation}
where $c_{H}^2=8$. We take the higssino mass $m_{\tilde{h}}= \mu = 500\,$GeV. 
The yield of the axino from the higgsino decay is estimated as
$Y_{\tilde{a}}^{(\tilde{h})}\simeq 5 \times 10^{-4} g_{\tilde{h}} M_{\rm P} 
\Gamma_{\tilde{h}}/m_{\tilde{h}}^2$, where $g_{\tilde{h}}=2$ is
the higgsino degrees of freedom~\cite{CKKR}. Then, the axino production from this decay 
is dominant at the low reheating temperature, $T_{\rm RH} \lesssim 5 \times 10^7$\,GeV. 
The density of the axino is given by \cite{Chun,Choi:2011yf}
\begin{equation}
\label{appOmegaDFSZ1}
\Omega_{\tilde{a}}^{\mathrm{TH} (\tilde{h})}h^2 
= 0.5 \left( \frac{m_{\tilde{a}}}{10\rm MeV}\right)
\left(\frac{f_a}{10^{12}\,\mathrm{GeV}}\right)^{-2}.
\end{equation}

\section*{Acknowledgments}
The work is supported by 
MEXT KAKENHI Grant Numbers 15H05889 (M.K.) and 25400248 (M.K.), 
and also by World Premier
International Research Center Initiative (WPI Initiative), MEXT, Japan.


\end{document}